\documentclass[11pt]{article}
\usepackage[a4paper,margin=2.25cm]{geometry}
\usepackage[T1]{fontenc}
\usepackage{lmodern}
\usepackage{amsmath,amssymb,mathtools,bm,mathrsfs,amsthm}
\usepackage{booktabs,longtable,array}
\usepackage{microtype}
\usepackage[hidelinks]{hyperref}
\allowdisplaybreaks
\newcommand{\hb}{\hbar}
\newcommand{\comm}[2]{\left[#1,#2\right]}
\newcommand{\acom}[2]{\left\{#1,#2\right\}_{+}}
\newcommand{\pb}[2]{\left\{#1,#2\right\}_{\rm P}}
\newcommand{\dd}{\mathrm d}
\newcommand{\PVI}{P_{\rm VI}}
\newcommand{\Id}{\mathcal I}
\newtheorem{theorem}{Theorem}

\hypersetup{pdftitle={Painleve VI superintegrability and Heun-Racah algebras},pdfauthor={Ian Marquette and Sarah Post}}
\title{Painlev\'e VI superintegrability and Heun--Racah algebras}

\author{Ian Marquette\\{\small Department of Mathematical and Physical Sciences, La Trobe University, Melbourne, Australia}\\{\small\texttt{i.marquette@latrobe.edu.au}}\\[0.7em]
Sarah Post\\{\small Department of Mathematics, University of Hawai\textquoteleft i at M\=anoa, Honolulu, USA}\\{\small\texttt{spost@hawaii.edu}}}

\date{26 October 2026}

\begin{document}
\maketitle

\begin{abstract}
We study the two fourth-order quantum superintegrable systems in the Euclidean plane that separate in polar coordinates and whose angular potentials are determined by the general four-parameter sixth Painlev\'e transcendent.  One system has an oscillator radial term and the other an extended-Coulomb radial term.  For both systems we give the fourth-order quantum integral explicitly, derive the cubic symmetry algebra and its Casimir, and construct a deformed-oscillator realization.  At fixed energy the two algebras are specializations of the Heun--Racah algebra.  Their structure functions contain the same quartic angular factor, determined by the Painlev\'e~VI parameters, while different radial factors distinguish oscillator and Coulomb dynamics.  The Coulomb integral is obtained by transforming the Cartesian Hermitian symmetry at the operator level.  The lower-order terms generated by the polar coordinate change are essential for commutation with the Hamiltonian.  The two quantum systems are also related by the Levi-Civita transformation and coupling-constant metamorphosis, which gives a direct correspondence between their Painlev\'e parameters, algebras, Casimirs and representations.  Earlier exceptional-Jacobi models arise at special rational parameter values, whereas the results here apply to the full four-parameter Painlev\'e~VI family.  We also compare the quantum results with the corresponding classical Poisson algebras.
\end{abstract}

\section{Introduction}
A two-dimensional quantum Hamiltonian $H$ is superintegrable when it admits two additional symmetry operators $X$ and $Y$ such that $H,X,Y$ are algebraically independent.  Their commutators often generate a polynomial algebra, and the Casimir and finite-dimensional representations of that algebra can provide spectral information without solving the Schr\"odinger equation directly \cite{MillerPostWinternitz,BonatsosDaskaloyannisKokkotas,Daskaloyannis,MarquetteQuartic}.  Painlev\'e transcendents entered this subject through the third-order systems separating in Cartesian coordinates \cite{GravelWinternitz2002,Gravel2004}, whose polynomial algebras and representations were worked out in \cite{MarquetteWinternitz2007,Marquette2009}; see \cite{PostWinternitz2015} for the general framework of $N$th-order integrals and \cite{MarquetteWinternitz2019} for a review of the Painlev\'e cases.

The fourth-order systems that separate in polar coordinates contain an especially rich family of exotic potentials \cite{ELW2017,ELWCorrection}, as do their Cartesian counterparts \cite{MarquetteSajediWinternitz2017}.  With
\begin{equation}
V(r,\theta)=R(r)+\frac{S(\theta)}{r^2},
\end{equation}
the compatibility conditions reduce the angular problem to the Cosgrove--Scoufis SD-I.a equation \cite{CosgroveScoufis}, whose general solution is expressed through the four-parameter sixth Painlev\'e transcendent.  Two radial choices are relevant here: the oscillator term $R(r)=\lambda r^2$ and the extended-Coulomb term $R(r)=a/r$.  Each system has a second-order separation integral and an independent fourth-order integral, but the radial dynamics and leading fourth-order tensors are different.  Painlev\'e~VI had already appeared in third-order polar superintegrability \cite{TremblayWinternitz2010}.  Related fourth-order standard potentials and the general $N$th-order polar setting are discussed in \cite{ELVWY2018,ERWY2018}.

In earlier work we studied rational members of the oscillator family using exceptional Jacobi polynomials \cite{GomezUllateKamranMilson2009,Quesne2008,MPR2020,MPR2022}.  Those special cases have a cubic quantum algebra and a deformed-oscillator realization, with structure constants expressed through the SD-I.a parameters.  Here we show that the same algebraic structure extends to a general Painlev\'e~VI angular potential: the oscillator algebra, Casimir and representation formulas hold for arbitrary values of the four Painlev\'e parameters.  We also carry out the corresponding calculation for the Coulomb branch, for which no four-parameter quantum polynomial algebra had previously been displayed.

For both systems we begin with the differential symmetry: we write the fourth-order integral in polar coordinates, establish its commutation with the Hamiltonian using the Painlev\'e determining equation, and only then derive the polynomial algebra.  The order matters in the Coulomb case, where the integral is classified in Cartesian form and its transformation to polar coordinates produces lower-order terms that have no counterpart in the principal symbol.  Once these terms are included, the oscillator and Coulomb calculations run in parallel.

At fixed energy the resulting cubic algebras are central specializations of the Heun--Racah algebra \cite{VinetZhedanovHeunHahn,BergeronEtAl}.  Their deformed-oscillator structure functions share the same quartic angular factor, which is expressed through the monodromy parameters of $P_{\rm VI}$, while the remaining factor carries the radial dynamics.  We also determine the finite-dimensional algebraic modules and compare their energy formulas with direct radial separation.  Which algebraic module is realized physically still depends on the chosen real angular solution and its boundary conditions.

The two quantum systems are related by the Levi-Civita transformation together with coupling-constant metamorphosis \cite{HGDR1984,PostWinternitz2010,KalninsMillerPost2010}.  We derive the two branches independently and then use this correspondence as a cross-check, relating their Painlev\'e parameters, structure constants, Casimirs and representations.  The classical Poisson algebras are discussed at the end. As the classical limit is singular, they serve for comparison and not as a source for the quantum coefficients.

Section~\ref{sec:framework} introduces the two Hamiltonians and the notation used in the calculations.  Sections~\ref{sec:oscop}--\ref{sec:oscspec} treat the oscillator symmetry, algebra and representations.  Sections~\ref{sec:cop}--\ref{sec:cspec} give the corresponding Coulomb analysis.  Section~\ref{sec:classical} compares the classical Poisson algebras, and Section~\ref{sec:unified} summarizes the relation between the two quantum systems.

\section{Two quantum polar models and notation}
\label{sec:framework}
The two quantum Hamiltonians studied in parallel are
\begin{align}
H_{\rm O}&=-\frac{\hb^2}{2}\left(\partial_r^2+\frac1r\partial_r+\frac1{r^2}\partial_\theta^2\right)
+\lambda r^2+\frac{T'(\theta)}{r^2},\qquad
X_{\rm O}=-\hb^2\partial_\theta^2+2T'(\theta),\label{eq:twoHO}\\
H_{\rm C}&=-\frac{\hb^2}{2}\left(\partial_r^2+\frac1r\partial_r+\frac1{r^2}\partial_\theta^2\right)
+\frac{a}{r}+\frac{T_{\rm C}'(\theta)}{r^2},\qquad
X_{\rm C}=-\hb^2\partial_\theta^2+2T_{\rm C}'(\theta).\label{eq:twoHC}
\end{align}
The two systems have the same form of polar separation and the same SD-I.a/$P_{\rm VI}$ angular equation, but their fourth-order tensors and radial terms are different.

The four-parameter Hamiltonians and fourth-order integrals were obtained in the polar classification of Escobar-Ruiz, L\'opez Vieyra and Winternitz \cite{ELW2017,ELWCorrection}, together with the corresponding classical Poisson algebras.  Table~\ref{tab:status} summarizes what was available and what is added here.

\begin{table}[tbp]
\centering
\caption{Status of the two quantum branches and their classical counterparts.}
\label{tab:status}
\small
\begin{tabular}{>{\raggedright\arraybackslash}p{3.1cm}>{\raggedright\arraybackslash}p{4.6cm}>{\raggedright\arraybackslash}p{6.8cm}}
\toprule
object & previously available & present paper\\
\midrule
quantum oscillator & four-parameter Hamiltonian and fourth-order integral, together with cubic algebras for rational exceptional-Jacobi specializations & four-parameter quantum closure, Heun--Racah identification, Casimir and deformed-oscillator representations\\
quantum Coulomb & four-parameter Hamiltonian and fourth-order integral & polar operator form, four-parameter quantum closure, Heun--Racah identification, Casimir and deformed-oscillator representations\\
classical systems & classical angular equations and Poisson algebras & comparison with the quantum structures, using the original classical conventions\\
\bottomrule
\end{tabular}
\end{table}

Here and below, ``generic'' means that no special rational Painlev\'e solution is chosen.  In the representation theory it also means that parameters are taken away from the special algebraic loci where roots collide or exceptional lattices occur.

We distinguish the three brackets used below.  For quantum operators we write
\begin{equation}
\comm{F}{G}=FG-GF,\qquad \acom{F}{G}=FG+GF,
\label{eq:bracketconventions}
\end{equation}
whereas for classical phase-space functions we use
\begin{align}
\pb{f}{g}={}&\partial_r f\,\partial_{p_r}g-\partial_{p_r}f\,\partial_rg
+\partial_\theta f\,\partial_{p_\theta}g-\partial_{p_\theta}f\,\partial_\theta g.
\label{eq:PBdefinition}
\end{align}
In the quantum sections the fifth-order generator is defined by \(C=\comm{A}{B}\).  The classical section follows the convention of \cite{ELW2017}, \(C^{\rm cl}=\pb{Y^{\rm cl}}{X^{\rm cl}}\), whose orientation is opposite to the quantum one.

For the quantum calculations we first place all derivatives in normal order and then use the angular equation to eliminate higher derivatives of the Painlev\'e function.  If $L=\sum_{m,n}F_{mn}\partial_r^m\partial_\theta^n$, the operator identity $L=0$ is then equivalent to the vanishing of the coefficient functions $F_{mn}$. These calculations were carried out symbolically in Maple and Mathematica, and no particular Painlev\'e solution was substituted.

Normal ordering uses the repeated Leibniz rule.  For two monomials one uses
\begin{align}
\bigl(p\,\partial_r^m\partial_\theta^n\bigr)
\bigl(q\,\partial_r^k\partial_\theta^\ell\bigr)
={}&\sum_{\alpha=0}^{m}\sum_{\beta=0}^{n}
\binom{m}{\alpha}\binom{n}{\beta}\,
 p\,\partial_r^\alpha\partial_\theta^\beta(q)\,
 \partial_r^{m-\alpha+k}\partial_\theta^{n-\beta+\ell}.
\label{eq:compositionrule}
\end{align}
The composition rule is applied before the angular differential relations are substituted.

For convenience, the longer calculations are first written at $\hbar=1$ and the powers of $\hbar$ are then restored by homogeneity.  For the oscillator we set \(T=\hbar^2\tau\) and \(\lambda=\hbar^2\lambda_0\).  For the Coulomb branch we set \(T_{\rm C}=\hbar^2\tau_{\rm C}\) and \(a=\hbar^2a_0\).  The constants $c_{11},c_{12},c_{31},c_{32}$ scale as $\hbar^2$ and $K_1,K_2$ as $\hbar^4$.  In both cases \(H=\hbar^2H_0\), \(X=\hbar^2X_0\), \(Y=\hbar^4Y_0\), and hence the fifth-order commutator generator scales as \(C=\hbar^6C_0\).  These weights determine the powers of \(\hbar\) appearing in the algebras and Casimirs.

For the confining branch of the polar classification whose leading fourth-order coefficient is $B_3$ in the notation of \cite{ELW2017} (their Case~II, here with $R(r)=\lambda r^2$), after a planar rotation and normalization of the leading fourth-order tensor, we use
\begin{equation}
H=-\frac{\hb^2}{2}\left(\partial_r^2+\frac1r\partial_r+\frac1{r^2}\partial_\theta^2\right)
+\lambda r^2+\frac{T'(\theta)}{r^2},
\label{eq:H}
\end{equation}
with second-order separation integral
\begin{equation}
X=-\hb^2\partial_\theta^2+2T'(\theta),
\qquad [H,X]=0.
\label{eq:X}
\end{equation}
The confining regime has $\lambda>0$.  The function $T$ is a local solution of the nonlinear compatibility equation imposed by the fourth-order symmetry.  No rational or algebraic specialization is assumed.

The fourth-order integral $Y$ is written in normal order in Section~\ref{sec:oscop}.  It is formally symmetric in the polar measure $r\,\dd r\,\dd\theta$ and satisfies
\begin{equation}
[H,Y]=0.
\label{eq:HYzero}
\end{equation}
The determining equations imply both properties for any solution $T$ of the angular equation.

The remaining determining equation is the fifth-order ODE
\begin{align}
0={}&\hb^2\left(16sT'-20cT''-10sT^{(3)}+\frac52cT^{(4)}+\frac14sT^{(5)}\right)
 +(16cT'+12sT''-2cT^{(3)})T\nonumber\\
&+24s(T')^2-
\left(24cT''+3sT^{(3)}+4(c_{11}c+c_{12}s)\right)T'
-3s(T'')^2\nonumber\\
&+3(c_{12}c-c_{11}s)T''
+\frac12(c_{11}c+c_{12}s)T^{(3)},
\qquad s=\sin2\theta,\quad c=\cos2\theta.
\label{eq:PVI5th}
\end{align}
Here $c_{11},c_{12}$ are constants of the lower-order determining equations, as in Eq.~(59) of \cite{ELW2017}.  The equation integrates twice, with integration constants $K_1,K_2$ (see \eqref{eq:jetflow}), and is transformed to the SD-I.a equation of Cosgrove and Scoufis \cite{CosgroveScoufis}.  In its standard second-degree second-order form,
\begin{align}
(\Sigma'')^2=-\frac4{f(x)^2}\Bigl[&q_2\Sigma'(x\Sigma'-\Sigma)^2+q_3(\Sigma')^2(x\Sigma'-\Sigma)
+q_7(\Sigma')^2\nonumber\\
&+q_8(x\Sigma'-\Sigma)+q_9\Sigma'+q_{10}\Bigr],
\label{eq:SDIa}
\end{align}
where $f(x)=x(x-1)$ and $q_2=-q_3=1$.  We write the SD-I.a function as $\Sigma(x)$, reserving $\mathsf W$ for the Heun--Racah generator introduced later.  The four constants $q_7,q_8,q_9,q_{10}$ parametrize the general $\PVI$ solution.

At $\hb=1$, the SD-I.a constants are
\begin{equation}
q_7=\frac{1-c_{12}}{16},\qquad
q_8=\frac{K_1}{64},\qquad
q_9=-\frac{2K_1-16K_2+c_{11}^2+c_{12}^2-4}{256},
\label{eq:qmap}
\end{equation}
while $q_{10}$ is the final first integral of the SD-I.a flow.  At arbitrary $\hb$ the dimensionless constants are
\begin{align}
q_7&=\frac{\hb^2-c_{12}}{16\hb^2},\qquad
q_8=\frac{K_1^{\rm phys}}{64\hb^4},\nonumber\\
q_9&=-\frac{2K_1^{\rm phys}-16K_2^{\rm phys}+c_{11}^2+c_{12}^2-4\hb^4}{256\hb^4}.
\label{eq:qmaphbar}
\end{align}
Here $K_j^{\rm phys}$ are the constants of \eqref{eq:jetflow} written for the physical $T$, $c_{11}$ and $c_{12}$.  They scale as $\hb^4$, and \eqref{eq:qmaphbar} reduces to \eqref{eq:qmap} at $\hb=1$.  This is the normalization associated with the choice $x=\sin^2\theta$ used in Section~\ref{sec:oscop}.  Equation \eqref{eq:PVI5th} is Eq.~(59) of \cite{ELW2017} with $B_3=1$.  Their integrated equation (62), written in $z=\tan2\theta$, agrees with \eqref{eq:jetflow} where $\cos2\theta>0$, up to a factor $8$, when their constants are $c_{11}/2$, $c_{12}/2$, $K_1/2$ and $K_2$.  Their parameter map (64) corresponds to the negative square root in their substitution (63), or equivalently to the reflected variable $x=\cos^2\theta$.  This changes $(q_8,q_9)$ into $(-q_8,q_9+q_8)$.  See Section~\ref{sec:cop}.
The primitive $T$ also has an additive gauge freedom that leaves the Hamiltonian unchanged.  The simultaneous change $T\mapsto T+\varkappa$, $c_{11}\mapsto c_{11}+4\varkappa$ and $K_2\mapsto K_2+\tfrac12c_{11}\varkappa+\varkappa^2$, with the original value of $c_{11}$ on the right, leaves the determining equations, the fourth-order integral and $q_7,\ldots,q_{10}$ unchanged, so the additive constant of $T$ is not a physical parameter.

\section{Oscillator branch: fourth-order quantum symmetry}
\label{sec:oscop}
This section gives the fourth-order oscillator integral in normal-ordered polar form and proves that it commutes with the Hamiltonian. The polynomial algebra it generates is the subject of Section~\ref{sec:oscalg}.

\paragraph{Fourth-order integral.}

Set
\begin{equation}
\tau=\frac{T}{\hb^2},\qquad a_{11}=\frac{c_{11}}{\hb^2},\qquad
a_{12}=\frac{c_{12}}{\hb^2},\qquad \lambda_0=\frac{\lambda}{\hb^2},
\qquad D=\partial_\theta,
\end{equation}
and write $s=\sin2\theta$, $c=\cos2\theta$, $\tau_j=D^j\tau$.  The fourth-order integral is
\begin{equation}
Y=\hb^4\left(\mathcal U_2\partial_r^2+\frac1r\mathcal U_1\partial_r+
\frac1{r^2}\mathcal U_0+\lambda_0r^2\mathcal U_\lambda\right),
\end{equation}
where every coefficient stands to the left of $D$ and
\begin{align}
\mathcal U_2={}&2cD^2-4sD-a_{11}s+a_{12}c+4\tau s-4\tau_1c-4c,\\
\mathcal U_1={}&-4sD^3-10cD^2
+2(-a_{11}c-a_{12}s+4\tau c+6\tau_1s+6s)D\nonumber\\
&+a_{11}s-a_{12}c-4\tau s+12\tau_1c+6\tau_2s+4c,\\
\mathcal U_0={}&-2cD^4+8sD^3+
(a_{11}s-a_{12}c-4\tau s+8\tau_1c+12c)D^2\nonumber\\
&+2(a_{11}c+a_{12}s-4\tau c-10\tau_1s+4\tau_2c-4s)D+F_0,\\
\mathcal U_\lambda={}&-4cD^2+8sD+2a_{11}s-2a_{12}c-8\tau s+8\tau_1c+8c,
\end{align}
with
\begin{align}
F_0=-\frac12\Bigl[&-4a_{11}\tau_1s+2a_{11}\tau_2c+4a_{12}\tau_1c+2a_{12}\tau_2s
+16\tau\tau_1s-8\tau\tau_2c\nonumber\\
&-16\tau_1^2c-12\tau_1\tau_2s+4\tau_2s+\tau_4s\Bigr].
\end{align}
A direct calculation gives
\begin{equation}
Y^{\dagger_{\rm formal}}=Y\quad\text{in the measure }r\,\dd r\,\dd\theta,
\qquad [H,Y]=0,
\end{equation}
using the nonlinear equation \eqref{eq:PVI5th} and its differential consequences.

\begin{theorem}[Quantum oscillator integral]
For every solution $T$ of \eqref{eq:PVI5th}, in particular for the unrestricted four-parameter $P_{\rm VI}$ family, the normal-ordered operator displayed above is formally symmetric in the polar measure and satisfies \(\comm{H}{Y}=0\) identically as a differential operator.
\end{theorem}
For an arbitrary function $T$, the commutator simplifies to
\begin{equation}
\comm{H}{Y}=\hb^2r^{-4}\acom{\mathcal E_5[T]}{\partial_\theta},
\label{eq:HYE5}
\end{equation}
where $\mathcal E_5[T]$ is the right-hand side of \eqref{eq:PVI5th}.  Hence $\comm{H}{Y}=0$ if and only if \eqref{eq:PVI5th} holds.

\paragraph{Using the angular equation.}

After the fifth-order equation is integrated, the angular function satisfies a third-order equation.  At a regular point, $T$, $T'$ and $T''$ may therefore be chosen as local initial data. The equation then fixes $T'''$, and differentiating it fixes all higher derivatives.  No explicit Painlev\'e~VI solution enters at any stage.

For the calculation set $\hb=1$, put $u=\tan\theta$, and write $\tau_0=T$, $\tau_1=T'$ and $\tau_2=T''$.  Then
\begin{equation}
s=\frac{2u}{1+u^2},\qquad c=\frac{1-u^2}{1+u^2},
\end{equation}
and the third-order relation gives
\begin{align}
s^2\tau_3={}&K_1c-8K_2-2s^2c_{12}\tau_1+6s^2\tau_1^2
+8(1+s^2)\tau_1-2cs\,c_{11}\tau_1\nonumber\\
&+8cs\tau_0\tau_1-4cs\tau_2+4c_{11}\tau_0-8\tau_0^2.
\label{eq:jetflow}
\end{align}
Along a solution, $du/d\theta=1+u^2$, $d\tau_0/d\theta=\tau_1$, $d\tau_1/d\theta=\tau_2$ and $d\tau_2/d\theta=\tau_3$, so the total derivative $d/d\theta$, written in the variables $(u,\tau_0,\tau_1,\tau_2)$, is
\begin{equation}
\mathscr D=(1+u^2)\partial_u+\tau_1\partial_{\tau_0}
+\tau_2\partial_{\tau_1}+\tau_3\partial_{\tau_2},
\end{equation}
with $\tau_3$ given by \eqref{eq:jetflow}.  Thus $\mathscr D$ is neither a new operator nor a new equation: $T^{(4)}=\mathscr D\tau_3$, $T^{(5)}=\mathscr D^2\tau_3$, and so on.  After these substitutions, every coefficient in the commutators is a rational expression in $r,u,T,T',T''$ and the constants, and the vanishing of those expressions proves the identities for every local solution of the angular equation.

The fourth SD-I.a constant can be handled in the same way.  The change of variables that brings the integrated angular equation to the SD-I.a form \eqref{eq:SDIa} is
\begin{equation}
x=\frac{1-c}{2},\qquad
\Sigma=\frac{s\tau_0}{4}-\frac{sc_{11}}{16}-\frac{c(1-c_{12}/2)}{8},
\qquad \partial_x=s^{-1}\mathscr D,
\end{equation}
where $\partial_x=s^{-1}\mathscr D$ on any interval where $s\neq0$, because $dx/d\theta=\sin2\theta=s$. Thus $\Sigma(x)$ is determined by $T$ and is not an independent function.  Writing $U=x\Sigma_x-\Sigma$ for the combination that recurs in \eqref{eq:SDIa}, and solving \eqref{eq:SDIa} for its last constant with \eqref{eq:qmap}, gives
\begin{align}
q_{10}=-\frac14\Bigl[&x^2(1-x)^2\Sigma_{xx}^2
+4\Sigma_xU^2-4\Sigma_x^2U
+4q_7\Sigma_x^2+4q_8U+4q_9\Sigma_x\Bigr].
\end{align}
Differentiating along the flow gives
\begin{equation}
\mathscr Dq_{10}=0,
\end{equation}
so $q_{10}$ is constant along every solution of the third-order equation: it is the fourth first integral, reconstructed from the local data $(u,T,T',T'')$.

The first three SD-I.a constants are fixed by $c_{12},K_1,K_2$, whereas the value of $q_{10}$, constant along each solution, varies from one solution to another and so depends on the initial data.  To see that the four constants are locally independent, fix a regular point and hold $T$, $T'$ and $c_{11}$ fixed, so that $T''=\tau_2$ supplies the remaining local freedom. Here $\tau_2$ is an initial value of the third-order equation, not a parameter of the Hamiltonian.  The Jacobian of $(c_{12},K_1,K_2,\tau_2)\mapsto(q_7,q_8,q_9,q_{10})$ is
\begin{equation}
\det\frac{\partial(q_7,q_8,q_9,q_{10})}{\partial(c_{12},K_1,K_2,\tau_2)}
=\frac{s^2\tau_2+2sc\tau_1-4\tau_0+c_{11}}{2^{23}},
\end{equation}
which is not identically zero.  Hence the four SD-I.a constants are locally independent for generic data, and the construction covers the full four-parameter $P_{\rm VI}$ family rather than a lower-dimensional specialization. A zero of this Jacobian at a particular point only means that these local coordinates degenerate there.

\section{Oscillator branch: polynomial algebra and Heun--Racah structure}
\label{sec:oscalg}
The second- and fourth-order integrals generate a cubic algebra, which is most transparent after the quantum shift
\begin{equation}
A=X-\hb^2,\qquad B=Y,\qquad C=[A,B].
\label{eq:ABC}
\end{equation}
The Hamiltonian $H$ is central.  Normal ordering and use of the SD-I.a relations give
\begin{align}
[A,B]&=C,\label{eq:alg1}\\
[A,C]&=\mathfrak b\acom{A}{B}+\mathfrak f,\label{eq:alg2}\\
[B,C]&=-\mathfrak b B^2+\mathfrak g A^3+\mathfrak hA^2+\mathfrak iA+\mathfrak j,
\label{eq:alg3}
\end{align}
with
\begin{align}
\mathfrak b&=8\hb^2,\qquad
\mathfrak f=-1024q_8\hb^6H,\label{eq:coeff1}\\
\mathfrak g&=-1024\lambda\hb^2,\label{eq:coeff2}\\
\mathfrak h&=384\hb^2H^2-1024\lambda\hb^4(10+12q_7),\label{eq:coeff3}\\
\mathfrak i&=2048\hb^4(1+2q_7)H^2
-1024\lambda\hb^6(24+64q_7-16q_8-32q_9),\label{eq:coeff4}\\
\mathfrak j&=2048\hb^6(1+4q_7-2q_8-4q_9)H^2\nonumber\\
&\qquad-16384\lambda\hb^8(1+4q_7-2q_8-4q_9+4q_{10}).
\label{eq:coeff5}
\end{align}
The calculation uses only the angular relations described in Section~\ref{sec:oscop}.

\begin{theorem}[Generic oscillator $P_{\rm VI}$ algebra]
For the Hamiltonian \eqref{eq:H}, the separation integral \eqref{eq:X} and the fourth-order integral of Section~\ref{sec:oscop}, the operators $H,A,B,C$ satisfy \eqref{eq:alg1}--\eqref{eq:coeff5} for every solution of the four-parameter SD-I.a/$P_{\rm VI}$ angular equation.
\end{theorem}

The generators have differential orders
\begin{equation}
\operatorname{ord}A=2,\qquad \operatorname{ord}B=4,\qquad \operatorname{ord}C=5.
\end{equation}
That $C$ has order exactly five follows from its principal symbol,
\begin{equation}
\pb{\sigma_2(X)}{\sigma_4(Y)}
=\frac{8p_\theta^3}{r^2}
\left(r^2p_r^2\sin2\theta+2rp_rp_\theta\cos2\theta-p_\theta^2\sin2\theta\right),
\label{eq:Csymbol}
\end{equation}
which is nonzero on a dense open set.

At fixed energy, the relations \eqref{eq:alg1}--\eqref{eq:alg3} are a specialization of the Heun--Racah algebra.  In the notation of Vinet and Zhedanov \cite{VinetZhedanovHeunHahn}, its generators $\mathsf Y,\mathsf W$ satisfy
\begin{align}
[\mathsf Y,[\mathsf Y,\mathsf W]]={}&g_1\mathsf Y^2+g_2\acom{\mathsf Y}{\mathsf W}
+g_3\mathsf Y+g_4\mathsf W+g_5\Id,\label{eq:HR1}\\
[\mathsf W,[\mathsf W,\mathsf Y]]={}&e_1\mathsf Y^2+e_2\mathsf Y^3+g_2\mathsf W^2
+g_1\acom{\mathsf Y}{\mathsf W}+g_3\mathsf W+g_6\mathsf Y+g_7\Id.
\label{eq:HR2}
\end{align}
The terms proportional to $e_1,e_2$ are the additional Heun terms beyond the ordinary Racah algebra, which is the symmetry algebra of the generic second-order superintegrable systems \cite{KalninsMillerPost2007,GenestVinetZhedanov2014}.  Here the coefficients are treated as free parameters.  In the realization of \cite{VinetZhedanovHeunHahn} they are specific functions of the Hahn data.

At fixed total energy $H=E$, set
\begin{equation}
\mathsf Y=A,\qquad \mathsf W=B.
\end{equation}
Since $[B,[B,A]]=-[B,C]$, comparison with \eqref{eq:alg2}--\eqref{eq:alg3} gives
\begin{align}
g_1&=0,& g_2&=\mathfrak b,&g_3&=0,&g_4&=0,&g_5&=\mathfrak f,\label{eq:HRmap1}\\
e_1&=-\mathfrak h,&e_2&=-\mathfrak g=1024\lambda\hb^2,&
 g_6&=-\mathfrak i,&g_7&=-\mathfrak j.
\label{eq:HRmap2}
\end{align}
Here the right-hand sides are evaluated at $H=E$. Before fixing the energy, the same statement holds over the central polynomial ring generated by $H$. In the notation of \cite{BergeronEtAl}, whose generators $(X,W,Z)$, with $Z=[W,X]$, are identified here with $(A,B,-C)$, the relations are their Eq.~(4.1) with $x_0=-\mathfrak f$, $x_1=x_2=x_3=x_4=0$, $x_5=-\mathfrak b$, $y_0=\mathfrak j$, $y_1=\mathfrak i$, $y_2=\mathfrak h$, $y_3=\mathfrak g$.

\begin{theorem}[Heun--Racah identification]
The oscillator symmetry algebra is a central-$H$ specialization of the abstract Heun--Racah algebra \eqref{eq:HR1}--\eqref{eq:HR2}, with $g_1=g_3=g_4=0$.  At each fixed energy it is an ordinary scalar-parameter Heun--Racah algebra with structure constants \eqref{eq:HRmap1}--\eqref{eq:HRmap2}.
\end{theorem}

In the confining regime $\lambda>0$,
\begin{equation}
e_2=1024\lambda\hb^2\neq0,
\end{equation}
and an affine change of generators $B\mapsto\alpha B+\beta A+\gamma$ alters the quadratic coefficient $e_1$ but only rescales $e_2$.  For $\lambda\neq0$ the cubic Heun term therefore cannot be removed, and the algebra lies in the Heun--Racah sector proper rather than in the Racah sector.  This is consistent with the relation between tridiagonalization, Heun equations and Racah-type algebras \cite{GrunbaumVinetZhedanov,BergeronEtAl}. A confluent (Laguerre) Heun algebra also occurs in the second-order Smorodinsky--Winternitz~II system \cite{CheaVinetZhedanov2026}.  We do not claim that the angular Schr\"odinger equation itself reduces to a Heun equation. The identification here is at the level of the symmetry algebra.

The Painlev\'e constants $q_7,q_8,q_9,q_{10}$ do not change the abstract algebra and they select a four-parameter central specialization inside the Heun--Racah class.  The coupling $\lambda$ controls the cubic Heun term, and the energy $E$ enters the remaining structure constants as a central parameter.

\paragraph{Heun--Racah parameters and the Racah limit.}
For later use it is convenient to display the identification in one table.  At fixed $H=E$ the Heun--Racah parameters are
\begin{center}
\begin{tabular}{cc}
\toprule
Heun--Racah parameter & polar $P_{\rm VI}$ value\\
\midrule
$g_1$ & $0$\\
$g_2$ & $8\hb^2$\\
$g_3$ & $0$\\
$g_4$ & $0$\\
$g_5$ & $-1024q_8\hb^6E$\\
$e_1$ & $-384\hb^2E^2+1024\lambda\hb^4(10+12q_7)$\\
$e_2$ & $1024\lambda\hb^2$\\
$g_6$ & $-2048\hb^4(1+2q_7)E^2+1024\lambda\hb^6(24+64q_7-16q_8-32q_9)$\\
$g_7$ & $-2048\hb^6(1+4q_7-2q_8-4q_9)E^2$\\
& $\quad+16384\lambda\hb^8(1+4q_7-2q_8-4q_9+4q_{10})$\\
\bottomrule
\end{tabular}
\end{center}
In the parametrization of \cite{BergeronEtAl} the cubic Heun coefficient is $y_3=\mathfrak g=-1024\lambda\hb^2$.  At $\lambda=0$ it vanishes, and the replacement $B\mapsto B+\eta A$ with $\eta=\pm4E$ also removes the remaining extra quadratic coefficient, so on the whole stratum $\lambda=0$ the fixed-energy algebra reduces to the Racah form.

\paragraph{Casimir operator.}
Ordering the products with the relations \eqref{eq:alg1}--\eqref{eq:alg3} gives the following Casimir:
\begin{align}
\mathcal K={}&C^2-8\hb^2\acom{A}{B^2}+64\hb^4B^2+2048q_8\hb^6HB
-512\lambda\hb^2 A^4\nonumber\\
&+\left[256\hb^2H^2-4096\lambda\hb^4(3+2q_7)\right]A^3\nonumber\\
&+\left[1024\hb^4(3+4q_7)H^2
-8192\lambda\hb^6(5+12q_7-2q_8-4q_9)\right]A^2\nonumber\\
&+\left[4096\hb^6(1+4q_7-2q_8-4q_9)H^2\right.\nonumber\\
&\qquad\left.-32768\lambda\hb^8(1+4q_7-2q_8-4q_9+4q_{10})\right]A.
\label{eq:Casimir}
\end{align}
When the differential operators are substituted, every noncentral term cancels and the Casimir collapses to
\begin{equation}
\mathcal K=-65536q_{10}\hb^8H^2+131072\lambda q_8^2\hb^{10}.
\label{eq:CasimirValue}
\end{equation}

The nonlinear term in the first adjoint relation is stable under the triangular changes of generator that preserve the filtration.  If
\begin{equation}
\widetilde B=\alpha B+q(A,H),\qquad \alpha\neq0,
\end{equation}
then $\widetilde C=[A,\widetilde B]=\alpha C$ and
\begin{equation}
[A,\widetilde C]
=\mathfrak b\acom{A}{\widetilde B}-2\mathfrak b A q(A,H)+\alpha\mathfrak f.
\end{equation}
The coefficient $\mathfrak b=8\hb^2$ is invariant in this class, so no polynomial triangular change of generators of this form reduces the algebra to a rank-one Lie or Weyl algebra.

\paragraph{Centrality and its value.}
To check centrality, write every word in the ordered form $A^iB^jC^k$, with $i,j,k\ge0$.  The only overlap ambiguity, $CBA$, is consistent with the defining relations.  Using $C=[A,B]$, relation \eqref{eq:alg2} moves $C$ through powers of $A$, while \eqref{eq:alg3} moves it through powers of $B$.  With the coefficients in \eqref{eq:Casimir}, both $[A,\mathcal K]$ and $[B,\mathcal K]$ vanish.

At fixed $H=E$, \eqref{eq:CasimirValue} becomes the scalar
\begin{equation}
\kappa(E)=-65536q_{10}\hb^8E^2+131072\lambda q_8^2\hb^{10},
\end{equation}
which enters the deformed-oscillator construction as one of the two finite-difference equations that determine the structure function.

\section{Oscillator branch: representations and spectrum}
\label{sec:oscspec}
The polynomial algebra admits a deformed-oscillator realization, which we construct following the standard approach for polynomial symmetry algebras \cite{BonatsosDaskaloyannisKokkotas,Daskaloyannis}, and then compare the resulting energy formulas with radial separation.
Let
\begin{equation}
[N,b^\dagger]=b^\dagger,\qquad [N,b]=-b,
\qquad b^\dagger b=\Phi(N),\qquad bb^\dagger=\Phi(N+1).
\label{eq:oscdef}
\end{equation}
Write $t=N+u$ and use
\begin{equation}
A=A(N),\qquad
B=d(N)+b^\dagger\rho(N)+\rho(N)b.
\label{eq:oscansatz}
\end{equation}
The $b^\dagger$ and $b$ parts of \eqref{eq:alg2} give the finite-difference condition
\begin{equation}
\bigl(A(N+1)-A(N)\bigr)^2
=\mathfrak b\bigl(A(N+1)+A(N)\bigr),
\label{eq:Adifference}
\end{equation}
whose nonconstant solution can be normalized as
\begin{equation}
A(t)=4\hb^2\left(t^2-\frac14\right),
\qquad
X(t)=A(t)+\hb^2=4\hb^2t^2.
\label{eq:AofN}
\end{equation}
The diagonal part of the same commutator fixes
\begin{equation}
d(t)=\frac{16q_8\hb^2E}{t^2-1/4}
\label{eq:dofN}
\end{equation}
when $H=E$.

The second cubic relation and the Casimir then determine the gauge-invariant product $\rho^2\Phi$.  A convenient oscillator gauge is
\begin{equation}
\rho(t)^2=\frac{1}{t(t+1)(2t+1)^2}.
\label{eq:rho}
\end{equation}
Define
\begin{equation}
\omega=\sqrt{2\lambda},\qquad \xi=2t-1.
\end{equation}
The structure function is
\begin{equation}
\Phi(t)=\hb^4\left(E^2-\hb^2\omega^2\xi^2\right)R_4(\xi^2),
\label{eq:Phi}
\end{equation}
where
\begin{equation}
R_4(w)=w^4+16q_7w^3-32(q_8+2q_9)w^2+256q_{10}w+256q_8^2.
\label{eq:R4}
\end{equation}
The quartic $R_4$ can also be read directly from the Jimbo--Miwa--Okamoto $\sigma$-form \cite{JimboMiwa1981,OkamotoPVI}.  For $f(x)=x(x-1)$, equation \eqref{eq:SDIa} is equivalent, wherever $\Sigma'\ne0$, to
\[
\Sigma'\bigl(x(x-1)\Sigma''\bigr)^2+\bigl[2\Sigma'(x\Sigma'-\Sigma)-\Sigma'^2-b_1b_2b_3b_4\bigr]^2=\prod_{k=1}^4(\Sigma'+b_k^2),
\]
with
\[
q_7=-\tfrac14e_1(b^2),\qquad q_8=-b_1b_2b_3b_4,\qquad q_9=-\tfrac14\bigl(e_2(b^2)-2b_1b_2b_3b_4\bigr),\qquad q_{10}=-\tfrac14e_3(b^2),
\]
where $e_k(b^2)$ are the elementary symmetric polynomials in $b_1^2,\ldots,b_4^2$.  It follows immediately that
\[
R_4(w)=\prod_{k=1}^4(w-4b_k^2).
\]
This identity is independent of the convention used to relate the $b_k$ to Painlev\'e~VI monodromy exponents \cite{OkamotoPVI}.

To make that relation explicit, write $P_{\rm VI}$ in its standard form \cite{DLMF32}, with time $x$, as
\begin{align*}
y''={}&\frac12\Bigl(\frac1y+\frac1{y-1}+\frac1{y-x}\Bigr)y'^2-\Bigl(\frac1x+\frac1{x-1}+\frac1{y-x}\Bigr)y'\\
&+\frac{y(y-1)(y-x)}{x^2(x-1)^2}\Bigl[\gamma_1+\gamma_2\frac{x}{y^2}+\gamma_3\frac{x-1}{(y-1)^2}+\gamma_4\frac{x(x-1)}{(y-x)^2}\Bigr],
\end{align*}
with
\begin{equation}
\gamma_1=\frac{(1-\theta_\infty)^2}{2},\qquad
\gamma_2=-\frac{\theta_0^2}{2},\qquad
\gamma_3=\frac{\theta_1^2}{2},\qquad
\gamma_4=\frac{1-\theta_x^2}{2}.
\label{eq:theta}
\end{equation}
Using the correspondence (51)--(54) of \cite{ELW2017}, with the branch $\sqrt{2\gamma_1}=1-\theta_\infty$, gives
\[
2b_{1,2}=\theta_0\pm\theta_1,\qquad 2b_{3,4}=\theta_x\pm\theta_\infty.
\]
Hence
\begin{align}
R_4(w)={}&\bigl[w-(\theta_0+\theta_1)^2\bigr]
\bigl[w-(\theta_0-\theta_1)^2\bigr]\nonumber\\
&\times\bigl[w-(\theta_\infty+\theta_x)^2\bigr]
\bigl[w-(\theta_\infty-\theta_x)^2\bigr].
\label{eq:R4factor}
\end{align}
The other branch $\sqrt{2\gamma_1}=\theta_\infty-1$ replaces $\theta_\infty$ by $2-\theta_\infty$.  The same pairing holds in the normalization of \cite{GIL2013}, Eq.~(2.17), whose exponents are half those of \eqref{eq:theta}.  Let
\begin{equation}
r_1=(\theta_0+\theta_1)^2,\quad
r_2=(\theta_0-\theta_1)^2,\quad
r_3=(\theta_\infty+\theta_x)^2,\quad
r_4=(\theta_\infty-\theta_x)^2,
\end{equation}
and let \(e_k(r)\) denote the elementary symmetric polynomials in \(r_1,\ldots,r_4\).  Comparing \eqref{eq:R4} and \eqref{eq:R4factor} gives the invariant dictionary
\begin{equation}
e_1(r)=-16q_7,\qquad
e_2(r)=-32(q_8+2q_9),\qquad
e_3(r)=-256q_{10},\qquad
e_4(r)=256q_8^2.
\label{eq:qthetaDictionary}
\end{equation}
The sign of \(q_8\) depends on the chosen SD-I.a/\(P_{\rm VI}\) branch, but the transition polynomial does not: under $(q_8,q_9)\mapsto(-q_8,q_9+q_8)$, both $q_8+2q_9$ and $q_8^2$ are unchanged.

Hence the structure function has ten linear transition roots in $\xi$:
\begin{equation}
\xi=\pm\frac{E}{\hb\sqrt{2\lambda}},\quad
\pm(\theta_0+\theta_1),\quad
\pm(\theta_0-\theta_1),\quad
\pm(\theta_\infty+\theta_x),\quad
\pm(\theta_\infty-\theta_x).
\label{eq:tenroots}
\end{equation}

\paragraph{Number-operator realization.}
To obtain \eqref{eq:AofN}, set
\begin{equation}
\Delta A_N=A(N+1)-A(N).
\end{equation}
The raising component of $C=[A,B]$, marked by the superscript $(+)$, is
\begin{equation}
C^{(+)}=b^\dagger\Delta A_N\rho(N),
\end{equation}
and a second commutator with $A$ gives
\begin{equation}
[A,C]^{(+)}=b^\dagger(\Delta A_N)^2\rho(N).
\end{equation}
On the right-hand side of \eqref{eq:alg2}, the raising component of the anticommutator is
\begin{equation}
\acom{A}{B}^{(+)}=b^\dagger\bigl(A(N+1)+A(N)\bigr)\rho(N).
\end{equation}
The central term $\mathfrak f$ has no raising component, hence \eqref{eq:Adifference}.  With $\mathfrak b=8\hb^2$ the quadratic ansatz $A(N)=a_2t^2+a_0$ yields
\begin{equation}
a_2=4\hb^2,\qquad a_0=-\hb^2,
\end{equation}
which gives \eqref{eq:AofN}.  The shift $u$ remains free until the endpoint conditions are imposed.

The diagonal component of \eqref{eq:alg2} reads
\begin{equation}
0=2\mathfrak b A(t)d(t)+\mathfrak f,
\end{equation}
so that
\begin{equation}
d(t)=-\frac{\mathfrak f}{2\mathfrak b A(t)}
=\frac{16q_8\hb^2E}{t^2-1/4}.
\end{equation}
The first Heun--Racah relation therefore fixes both the number-operator spectrum and the diagonal part of the fourth-order integral.

\paragraph{Transition amplitudes.}
Let
\begin{equation}
\mathcal A(t)=\rho(t-1)^2\Phi(t).
\end{equation}
The diagonal part of \eqref{eq:alg3} is linear in $\mathcal A(t)$ and $\mathcal A(t+1)$.  The diagonal part of the Casimir relation \eqref{eq:Casimir} gives a second linear equation for the same two quantities.  Solving the pair eliminates the oscillator gauge and gives
\begin{equation}
\mathcal A(t)=\frac{\hb^4\bigl(E^2-\hb^2\omega^2\xi^2\bigr)R_4(\xi^2)}{t(t-1)\xi^2},
\qquad \xi=2t-1.
\label{eq:transition}
\end{equation}
Choosing the gauge \eqref{eq:rho} converts \eqref{eq:transition} into the structure function \eqref{eq:Phi}.  Away from $t\in\{-1,0,1\}$, where the two diagonal equations become degenerate, this is the unique transition product compatible with the cubic relation and the Casimir value.

\paragraph{Tridiagonal representation.}
Let $|n\rangle$, $n=0,\ldots,p$, be an oscillator basis with
\begin{equation}
N|n\rangle=n|n\rangle,\qquad
b^\dagger|n\rangle=\sqrt{\Phi(n+1)}\,|n+1\rangle,\qquad
b|n\rangle=\sqrt{\Phi(n)}\,|n-1\rangle.
\end{equation}
Writing $t_n=n+u$, the generators act as
\begin{align}
A|n\rangle={}&4\hb^2\left(t_n^2-\frac14\right)|n\rangle,\\
B|n\rangle={}&d(t_n)|n\rangle
+\rho(t_n)\sqrt{\Phi(n+1)}\,|n+1\rangle
+\rho(t_n-1)\sqrt{\Phi(n)}\,|n-1\rangle,\\
C|n\rangle={}&\bigl(A(t_n+1)-A(t_n)\bigr)\rho(t_n)\sqrt{\Phi(n+1)}\,|n+1\rangle\nonumber\\
&-\bigl(A(t_n)-A(t_n-1)\bigr)\rho(t_n-1)\sqrt{\Phi(n)}\,|n-1\rangle.
\end{align}
$A$ is diagonal, while $B$ and $C$ are tridiagonal.  The matrix dimension $p+1$ is the algebraic degeneracy of one irreducible fixed-energy module, before accidental coincidences between different endpoint choices are considered.

\paragraph{Radial separation.}
Polar separation gives the same energy normalization and degeneracy directly.  Since
\begin{equation}
H=-\frac{\hb^2}{2}\left(\partial_r^2+\frac1r\partial_r\right)+\lambda r^2+\frac{X}{2r^2},
\label{eq:Hseparated}
\end{equation}
an $X$-eigenstate with eigenvalue $x$ leads to the radial equation
\begin{equation}
\left[-\frac{\hb^2}{2}\left(\frac{\dd^2}{\dd r^2}+\frac1r\frac{\dd}{\dd r}\right)
+\lambda r^2+\frac{x}{2r^2}\right]R(r)=ER(r).
\label{eq:radial}
\end{equation}
For $\lambda>0$ set $\omega=\sqrt{2\lambda}$.  The regular singular-oscillator radial spectrum has the algebraic form
\begin{equation}
E=\hb\omega\left(2n_r+1+\frac{\sqrt{x}}{\hb}\right),
\qquad n_r=0,1,2,\ldots,
\label{eq:radialenergy}
\end{equation}
for the branch with $\sqrt{x}/\hb$ chosen consistently with the angular boundary condition.

The deformed oscillator gives
\begin{equation}
X_n=4\hb^2(n+u)^2.
\label{eq:Xn}
\end{equation}
If the vacuum is placed at the angular transition root $\xi_0=\mu$, then
\begin{equation}
u=\frac{\mu+1}{2},\qquad
\frac{\sqrt{X_n}}{\hb}=\mu+2n+1
\label{eq:Xroot}
\end{equation}
for the positive branch.  Substituting \eqref{eq:Xroot} into \eqref{eq:radialenergy} gives
\begin{equation}
E=\hb\omega\bigl[\mu+2(n+n_r+1)\bigr].
\end{equation}
Hence a fixed-energy multiplet is obtained by holding
\begin{equation}
p=n+n_r
\label{eq:pshell}
\end{equation}
constant.  The states
\begin{equation}
(n,n_r)=(0,p),(1,p-1),\ldots,(p,0)
\label{eq:shellstates}
\end{equation}
all have
\begin{equation}
E=\hb\omega\,[\mu+2(p+1)],
\label{eq:separationenergy}
\end{equation}
which is the positive branch selected by the corresponding zero of the structure function.  The $(p+1)$-dimensional deformed-oscillator module therefore has a direct separated interpretation: the polynomial symmetry transfers one unit of angular algebraic excitation while lowering the radial quantum number by one, leaving the total energy unchanged.

The agreement fixes the $\hb$ and $\sqrt{2\lambda}$ normalization of the algebraic energy formula.  It also shows that a finite-dimensional symmetry module does not mean that either separated one-dimensional problem has a finite spectrum: the finite module is the fixed-total-energy shell inside the product of two infinite separated spectra.

\paragraph{Finite-dimensional modules.}
\label{sec:energy}
A module is called unitary if it carries a positive-definite inner product for which $A$ and $B$ are self-adjoint.  In the gauge $\rho\equiv1$, where $\Phi(n)=\mathcal A(t_n)$, the deformed-oscillator conditions for a $(p+1)$-dimensional unitary module are
\begin{equation}
 \Phi(0)=0,\qquad \Phi(p+1)=0,\qquad
 \Phi(n)>0\quad (1\leq n\leq p).
 \label{eq:finiteconditions}
\end{equation}
The factorization \eqref{eq:Phi}--\eqref{eq:tenroots} reduces these conditions to the ordering of the transition roots along the lattice.  Put
\begin{equation}
 \xi_n=2(n+u)-1=\xi_0+2n,
 \qquad P=p+1,
 \label{eq:lattice}
\end{equation}
and introduce the four nonnegative angular moduli
\begin{equation}
 a_1=|\theta_0+\theta_1|,\quad
 a_2=|\theta_0-\theta_1|,\quad
 a_3=|\theta_\infty+\theta_x|,\quad
 a_4=|\theta_\infty-\theta_x|,
 \label{eq:amoduli}
\end{equation}
when the monodromy combinations are real.  Also set
\begin{equation}
 e=\frac{|E|}{\hb\sqrt{2\lambda}}.
\end{equation}
Unitarity concerns the gauge-invariant entries $B_{n-1,n}^2=\mathcal A(t_n)$ of \eqref{eq:transition}. Since $t(t-1)=(\xi^2-1)/4$,
\begin{equation}
 \mathcal A(t_n)=\frac{8\lambda\hb^6\,(e^2-\xi_n^2)\prod_{j=1}^4(\xi_n^2-a_j^2)}{\xi_n^2(\xi_n^2-1)} .
 \label{eq:signcriterion}
\end{equation}
The gauge \eqref{eq:rho} has $\rho(t_n-1)^2=4/[\xi_n^2(\xi_n^2-1)]$, which is negative for $|\xi_n|<1$.  In that interval $\Phi(n)>0$ does not make $B$ Hermitian. In the gauge $\rho\equiv1$ one has $\Phi(n)=\mathcal A(t_n)$.
This gives the following necessary and sufficient algebraic criterion.

\begin{theorem}[Finite-dimensional unitary root-lattice criterion]
\label{thm:lattice}
Choose two transition roots $r_{\rm L},r_{\rm R}$ from
\begin{equation}
 \mathscr R=\{\pm e,\ \pm a_1,\ldots,\pm a_4\}
\end{equation}
with
\begin{equation}
 r_{\rm R}-r_{\rm L}=2P,
 \qquad P\in\mathbb N.
 \label{eq:rootspacing}
\end{equation}
Set $\xi_0=r_{\rm L}$ and $u=(r_{\rm L}+1)/2$, and assume that no lattice point $\xi_n$, $0\le n\le P$, lies in $\{-3,-2,-1,0,1\}$.  This holds whenever $r_{\rm L}\notin\mathbb Z$. Then the associated $P$-dimensional deformed-oscillator module with $H=E$ and Casimir value \eqref{eq:CasimirValue} exists, with $\mathcal A(t_n)$ given by \eqref{eq:transition}, and it is unitary if and only if
\begin{equation}
 \frac{(e^2-\xi_n^2)\prod_{j=1}^4(\xi_n^2-a_j^2)}{\xi_n^2-1}>0,
 \qquad n=1,\ldots,P-1.
 \label{eq:allpositivity}
\end{equation}
It is realized in the gauge $\rho\equiv1$, where $\Phi(n)=\mathcal A(t_n)$. The condition depends only on the transition roots.
\end{theorem}

For lattices avoiding $\{-3,-2,-1,0,1\}$ the theorem gives the complete classification of irreducible unitary modules on which $H$ acts as $E$ and $\mathcal K$ acts by \eqref{eq:CasimirValue}: every such module is obtained by selecting two roots with the required spacing and imposing the finite set of positivity inequalities.  Indeed, unitarity makes $A$ diagonalizable, and \eqref{eq:alg2} places its eigenvalues on a chain $A(t_n)$, $t_n=u+n$.  This chain cannot fold back on a lattice avoiding $\{-3,-2,-1,0,1\}$: since $A(t)=A(-t)$ pairs $\xi$ with $-2-\xi$, a step-two chain containing both would meet $-1$ or $\{-2,0\}$.  The diagonal part of \eqref{eq:alg2}, together with the two diagonal equations, whose determinant is proportional to $t(t-1)(t+1)$, makes the diagonal blocks of $B$ and the products of its off-diagonal blocks scalar, so every eigenvalue of $A$ on an irreducible module is simple.  For $P\le5$, elimination of the module equations shows that tridiagonal modules on lattices meeting $\{-3,-2,-1,0,1\}$ exist only on proper subvarieties of the parameter space, for example $q_8=0$, $R_4(1)=0$ or $R_4(k^2)=0$ with $k\in\mathbb Z$.  The three endpoint types have different spectral interpretations and are discussed separately below.

\paragraph{Angular--radial branches.}
Let the lower endpoint be an angular root $\mu\in\{\pm a_1,\ldots,\pm a_4\}$ and terminate on a radial root.  From
\begin{equation}
 \pm\frac{E}{\hb\sqrt{2\lambda}}-\mu=2P
\end{equation}
we obtain the sixteen formal energy families
\begin{equation}
 E_{\mu,\varepsilon}^{(P)}
 =\varepsilon\hb\sqrt{2\lambda}\,(\mu+2P),
 \qquad \varepsilon=\pm1.
 \label{eq:mixedenergies}
\end{equation}
The automorphism $(H,B,C)\mapsto(-H,-B,-C)$ of \eqref{eq:alg1}--\eqref{eq:alg3} leaves $\mathcal A$ invariant and maps $E^{(P)}_{\mu,\varepsilon}$ to $E^{(P)}_{\mu,-\varepsilon}$. Thus the sixteen oriented choices reduce to eight formal angular-root families $|E|=\hb\sqrt{2\lambda}\,|\mu+2P|$, $\mu\in\{\pm a_1,\ldots,\pm a_4\}$, before reality and positivity are imposed.

For the physical confining oscillator one takes $E>0$.  Consider the branch that starts at the largest angular modulus,
\begin{equation}
 a_{\max}=\max(a_1,a_2,a_3,a_4),
 \qquad \xi_0=a_{\max},
 \label{eq:amax}
\end{equation}
and ends at $+e$.  Then
\begin{equation}
 E=\hb\sqrt{2\lambda}\,[a_{\max}+2P],
 \label{eq:robustenergy}
\end{equation}
and every interior point satisfies
\begin{equation}
 a_j^2<\xi_n^2<e^2,
 \qquad j=1,\ldots,4,
 \quad n=1,\ldots,P-1.
\end{equation}
All five factors in the numerator of \eqref{eq:signcriterion} are therefore positive, and so is $\xi_n^2(\xi_n^2-1)$ because $\xi_n\ge2$.  This branch is algebraically unitary for real monodromy data away from the exceptional loci described above.  Whether it occurs in the Hilbert space is a separate analytic question: it does when the normalizable angular ground state of $X$ has eigenvalue $x$ with $\sqrt{x}/\hb=a_{\max}+1$, and which angular root is realized depends on the boundary behavior of the chosen $P_{\rm VI}$ branch.

Other angular--radial branches are governed by the same positivity condition \eqref{eq:allpositivity}.  If the starting angular root is not outermost, the lattice can cross one or more of the other squared angular roots, and the sign of \eqref{eq:signcriterion} alternates at each crossing of a squared root or of $\xi^2=1$.

The radial separation calculation above selects the positive-energy orientation and identifies
\begin{equation}
 P=p+1=n+n_r+1.
\end{equation}
When the angular ground state corresponds to $\xi_0=a_{\max}$, the module \eqref{eq:robustenergy} coincides with the full fixed-energy shell of $P$ states.  More generally, $a_{\max}$ is replaced by the angular root selected by the angular spectral problem.

\paragraph{Radial--radial branches.}
Choosing the two radial roots $-e,+e$ as endpoints imposes
\begin{equation}
 2e=2P,
 \qquad |E|=\hb\sqrt{2\lambda}\,P.
 \label{eq:rrenergy}
\end{equation}
This condition contains no $P_{\rm VI}$ parameter, and the lattice $\xi_n=-P+2n$ is integral. For even $P$ it contains $\xi_n=-2$ and $\xi_n=0$ with $n\le P-1$, where $A(t_n)=0$, so the diagonal part of \eqref{eq:alg2} forces $\mathfrak f=0$, i.e.\ $q_8E=0$.  For odd $P$ it passes through $t\in\{-1,0,1\}$, where the diagonal relations do not determine $\mathcal A$.  For $P=1,3,5$ the module equations require $R_4(1)=0$. Radial--radial modules therefore occur only on special parameter values.

\paragraph{Angular--angular branches.}
If both endpoints are angular roots $\mu,\nu$, then
\begin{equation}
 \nu-\mu=2P.
 \label{eq:aaresonance}
\end{equation}
This is a codimension-one arithmetic resonance among the four monodromy combinations.  The energy remains a free central parameter at the endpoint level. Unitarity further requires
\begin{equation}
 \frac{(e^2-\xi_n^2)\prod_{j=1}^4(\xi_n^2-a_j^2)}{\xi_n^2-1}>0
\end{equation}
throughout the interior.  Such modules occur only at special $P_{\rm VI}$ parameter values and are not part of the continuous-parameter spectrum considered here.

\paragraph{Degenerate roots and resonant cases.}
The classification above assumes simple transition roots.  Root collisions occur when two angular moduli coincide, when an angular root meets a radial root, or when one of the monodromy combinations vanishes.  At these special parameter values the representation can change qualitatively.  A double endpoint changes the order of the zero of $\Phi$ and can split, shorten or render reducible the corresponding oscillator module.  We record these collision points as boundaries of the unitary domains rather than analyze each special case separately.

Likewise, two angular--radial towers labeled by $\mu$ and $\mu'$ have coincident energies only if
\begin{equation}
 \mu-\mu'=2\ell,
 \qquad \ell\in\mathbb Z,
 \label{eq:towercoincidence}
\end{equation}
with a compensating shift in $P$.  Away from these arithmetic resonances, the intrinsic degeneracy of a module is exactly its oscillator dimension, with no extra accidental coincidence between distinct root choices.

\paragraph{From algebraic modules to the angular spectral problem.}
The inequalities above are necessary for a module to occur in the Schr\"odinger spectrum but are not sufficient.  One also needs a real Painlev\'e~VI branch on the chosen angular interval, an admissible treatment of any poles, and a self-adjoint realization of
\begin{equation}
 X=-\hb^2\partial_\theta^2+2T'(\theta).
\end{equation}
The radial problem selects the positive-energy branch and the regular radial solution.  At a real movable pole $\theta_*$, the local behavior $\tau=-(\theta-\theta_*)^{-1}+O(1)$ gives $2T'\simeq2\hb^2(\theta-\theta_*)^{-2}$, so the angular operator is limit point there and the physical problem is posed on the intervals between consecutive real poles.

\section{Coulomb branch: fourth-order quantum symmetry}
\label{sec:cop}
For the Coulomb branch the classified fourth-order integral is given in Cartesian form.  To compare it with the polar Hamiltonian and separation operator, we first transform the full Hermitian operator to polar coordinates.  The Levi-Civita relation in Section~\ref{sec:unified} later connects the two branches.

The branch with leading coefficient $A_2$ (Case~I of \cite{ELW2017}, here with $R(r)=a/r$) gives the second quantum family
\begin{equation}
 H_{\rm C}=-\frac{\hb^2}{2}\Delta+\frac{a}{r}+\frac{T_{\rm C}'(\theta)}{r^2},
 \qquad
 X_{\rm C}=-\hb^2\partial_\theta^2+2T_{\rm C}'(\theta),
 \label{eq:CoulombH}
\end{equation}
where $a<0$ is the attractive bound-state regime.  Its fifth-order quantum determining equation is
\begin{align}
0={}&\hb^2\left(-4\sin\theta\,T_{\rm C}'+8\cos\theta\,T_{\rm C}''+6\sin\theta\,T_{\rm C}^{(3)}
-\frac52\cos\theta\,T_{\rm C}^{(4)}-\frac12\sin\theta\,T_{\rm C}^{(5)}\right)\nonumber\\
&+\left(-4\cos\theta\,T_{\rm C}'-6\sin\theta\,T_{\rm C}''+2\cos\theta\,T_{\rm C}^{(3)}\right)T_{\rm C}
-12\sin\theta\,(T_{\rm C}')^2\nonumber\\
&+\left(24\cos\theta\,T_{\rm C}''+6\sin\theta\,T_{\rm C}^{(3)}
-8c_{31}\cos\theta-8c_{32}\sin\theta\right)T_{\rm C}'
+6\sin\theta\,(T_{\rm C}'')^2\nonumber\\
&+12(c_{32}\cos\theta-c_{31}\sin\theta)T_{\rm C}''
+4(c_{31}\cos\theta+c_{32}\sin\theta)T_{\rm C}^{(3)}.
\label{eq:Coulomb5th}
\end{align}
Here $c_{31},c_{32}$ are constants of the lower-order determining equations and their relation to \cite{ELW2017} is stated below.  The radial coupling $a$ cancels from this equation, so the Painlev\'e equation is entirely angular, while $a$ enters the radial equation and the lower-order terms of the fourth-order symmetry.

\paragraph{Using the angular equation.}
As in the oscillator case, \eqref{eq:Coulomb5th} integrates twice to a third-order relation, with integration constants $K_1,K_2$.  At a regular point, $T_{\rm C}$, $T_{\rm C}'$ and $T_{\rm C}''$ are local initial data, while the third-order relation determines $T_{\rm C}'''$ and hence all higher derivatives.  Set $\hb=1$, put
\begin{equation}
u=\tan\frac\theta2,\qquad
s=\sin\theta=\frac{2u}{1+u^2},\qquad
c=\cos\theta=\frac{1-u^2}{1+u^2},
\end{equation}
and write $\tau_0=T_{\rm C}$, $\tau_1=T_{\rm C}'$, $\tau_2=T_{\rm C}''$.  The third-order relation is
\begin{equation}
\tau_3=-\frac{\mathcal N_{\rm C}}{4u^2},
\label{eq:Cjet3}
\end{equation}
where
\begin{align}
\mathcal N_{\rm C}={}&-4K_1u^4+4K_1-K_2u^4-2K_2u^2-K_2
+2\tau_0^2u^4+4\tau_0^2u^2+2\tau_0^2\nonumber\\
&+8\tau_0\tau_1u^3-8\tau_0\tau_1u
+8c_{31}\tau_0u^4+16c_{31}\tau_0u^2+8c_{31}\tau_0
-24\tau_1^2u^2\nonumber\\
&+16c_{31}\tau_1u^3-16c_{31}\tau_1u-32c_{32}\tau_1u^2
-2\tau_1u^4-20\tau_1u^2-2\tau_1
-4\tau_2u^3+4\tau_2u.
\end{align}
As in the oscillator case, the total derivative $d/d\theta$ along a solution, written in the variables $(u,\tau_0,\tau_1,\tau_2)$, is
\begin{equation}
\mathscr D_{\rm C}=\frac{1+u^2}{2}\partial_u
+\tau_1\partial_{\tau_0}+\tau_2\partial_{\tau_1}
+\tau_3\partial_{\tau_2},
\label{eq:Cderivation}
\end{equation}
where now $du/d\theta=(1+u^2)/2$ because $u=\tan(\theta/2)$.  Then $T_{\rm C}^{(4)}=\mathscr D_{\rm C}\tau_3$, and all later derivatives follow recursively.

\paragraph{Painlev\'e~VI parameters.}
The third-order relation \eqref{eq:Cjet3} reduces to the same SD-I.a equation \eqref{eq:SDIa}, and the choice of branch in the transformation to $\Sigma(x)$ fixes the signs in the parameter map.  The constants $c_{31},c_{32},K_1,K_2$ are those of the integrated equation (47) of \cite{ELW2017}, where $c_{31},c_{32},K_1$ were rescaled by $2A_2$ ($A_2=1$) before integrating, and relative to their fifth-order equation (45), $c_{31}$ and $c_{32}$ are therefore halved.  With this convention \eqref{eq:Cjet3} coincides with their Eq.~(47), and the invariant map is
\begin{align}
 q_7^{\rm C}&=\frac{5\hb^2+16c_{32}}{16\hb^2},&
 q_8^{\rm C}&=-\frac{K_1}{\hb^4},\nonumber\\
 q_9^{\rm C}&=\frac{32K_1-8K_2-3\hb^4-64c_{31}^2
 -32\hb^2c_{32}-64c_{32}^2}{64\hb^4},
 \label{eq:Cqmap}
\end{align}
with $q_{10}^{\rm C}$ the fourth SD-I.a first integral.  The canonical transformation to \eqref{eq:SDIa} is
\begin{equation}
x=\frac{1-\cos\theta}{2},\qquad
\Sigma_{\rm C}=\frac{\sin\theta}{2\hb^2}\bigl(T_{\rm C}+2c_{31}\bigr)-\frac{(3\hb^2+8c_{32})\cos\theta}{8\hb^2},\qquad
\partial_x=\frac{2}{\sin\theta}\,\partial_\theta ,
\label{eq:CWmap}
\end{equation}
as in \cite{ELW2017}.  With $U_{\rm C}=x\partial_x\Sigma_{\rm C}-\Sigma_{\rm C}$, the fourth constant $q_{10}^{\rm C}$ is obtained from the same formula as $q_{10}$, with $(\Sigma,U,q_7,q_8,q_9)$ replaced by $(\Sigma_{\rm C},U_{\rm C},q_7^{\rm C},q_8^{\rm C},q_9^{\rm C})$.

The substitution in \cite{ELW2017} contains $\sqrt{x(1-x)}$ and therefore has two branches.  We choose the branch $x=(1-\cos\theta)/2$ on $0<\theta<\pi$, which gives \eqref{eq:CWmap}.  The reflected branch has $x=(1+\cos\theta)/2$ and $\Sigma_{\rm C}\mapsto-\Sigma_{\rm C}$.  On that branch
\[
q_8^{\rm C}=\frac{K_1}{\hb^4},\qquad
q_9^{\rm C}=-\frac{32K_1+8K_2+3\hb^4+64c_{31}^2+32\hb^2c_{32}+64c_{32}^2}{64\hb^4},
\]
which is the parameter map printed in \cite{ELW2017}.  For the branch \eqref{eq:CWmap}, direct differentiation of \eqref{eq:Cqmap} along the flow \eqref{eq:Cderivation} gives
\begin{equation}
\mathscr D_{\rm C}q_{10}^{\rm C}=0,
\label{eq:Cq10invariant}
\end{equation}
which fixes the convention used below.

At $\hb=1$ the same independence test as in the oscillator case, with $T_{\rm C}$, $T_{\rm C}'$ and $c_{31}$ held fixed at a regular point and the initial value $T_{\rm C}''=\tau_2$ as the remaining local datum, gives
\begin{equation}
\det\frac{\partial(q_7^{\rm C},q_8^{\rm C},q_9^{\rm C},q_{10}^{\rm C})}{\partial(c_{32},K_1,K_2,\tau_2)}
=\frac{2c_{31}+\tau_0-\sin\theta\cos\theta\,\tau_1-\sin^2\theta\,\tau_2}{64},
\label{eq:CJacobian}
\end{equation}
which is not identically zero, so the Coulomb branch also covers a full four-parameter SD-I.a family.  As on the oscillator side, the primitive has an additive gauge.  The simultaneous change $T_{\rm C}\mapsto T_{\rm C}+\varkappa$, $c_{31}\mapsto c_{31}-\varkappa/2$ and $K_2\mapsto K_2+8c_{31}\varkappa-2\varkappa^2$, with the original value of $c_{31}$ on the right, leaves the determining equations, the integral and the four SD-I.a invariants unchanged.

For later use we abbreviate the scalar combination
\begin{equation}
\Delta_{\rm C}=1+4q_7^{\rm C}-2q_8^{\rm C}-4q_9^{\rm C},
\label{eq:DC}
\end{equation}
not to be confused with the derivation $\mathscr D_{\rm C}$.  The quartic \eqref{eq:R4} with $q_k$ replaced by $q_k^{\rm C}$, still denoted $R_4$, will govern the angular transition factors of the Coulomb representation.

\paragraph{Fourth-order integral in polar coordinates.}
The fourth-order symmetry is originally written in Cartesian Hermitian form.  Passing to polar coordinates changes not only the principal symbol but also the lower-order terms.  For arbitrary coefficient functions $G_1,G_2,G_3$, normal ordering the Cartesian Hermitian quadratic block in polar derivatives gives
\begin{align}
\mathcal Q_{\rm Cart}={}&\acom{G_1}{\partial_r^2}+\acom{G_3}{\partial_r\partial_\theta}
+\acom{G_2}{\partial_\theta^2}
+\frac{2G_1}{r}\partial_r+\frac{G_3}{r}\partial_\theta\nonumber\\
&-\frac{r^2\partial_rG_2+2rG_2-2\partial_rG_1-2\partial_\theta G_3}{r}.
\label{eq:polarorderingidentity}
\end{align}
The last line consists of lower-order coordinate terms with no counterpart in the principal symbol.

For the $A_2$ branch $G_1=0$ and
\begin{align}
G_2={}&\frac{4\cos\theta\,T_{\rm C}'+2c_{32}\cos\theta-(T_{\rm C}+2c_{31})\sin\theta}{r}
 +a\cos\theta,\\
G_3={}&3\sin\theta\,T_{\rm C}'+(T_{\rm C}+2c_{31})\cos\theta+2c_{32}\sin\theta,
\end{align}
while
\begin{align}
G_4={}&\frac{\hb^2}{2r}\left(\sin\theta\,T_{\rm C}^{(4)}+4\cos\theta\,T_{\rm C}^{(3)}
-3\sin\theta\,T_{\rm C}''-2\cos\theta\,T_{\rm C}'\right)\nonumber\\
&-\frac2r\left(3\sin\theta\,T_{\rm C}'+\cos\theta\,T_{\rm C}+2c_{31}\cos\theta+2c_{32}\sin\theta\right)T_{\rm C}''\nonumber\\
&-2a\sin\theta\,T_{\rm C}+4a\cos\theta\,T_{\rm C}'
-4a(c_{31}\sin\theta-c_{32}\cos\theta)-a\hb^2\cos\theta.
\label{eq:CG4source}
\end{align}
The polar operator is
\begin{align}
Y_{\rm C}={}&\hb^4\acom{\partial_\theta^3}{\sin\theta\,\partial_r}
+\frac{\hb^4}{r}\acom{\partial_\theta^3}{\cos\theta\,\partial_\theta}\nonumber\\
&-\hb^2\left(\acom{G_3}{\partial_r\partial_\theta}+\acom{G_2}{\partial_\theta^2}
+\frac{G_3}{r}\partial_\theta
-\frac{r^2\partial_rG_2+2rG_2-2\partial_\theta G_3}{r}\right)+G_4.
\label{eq:CYcorrected}
\end{align}

\begin{theorem}[Quantum Coulomb integral]
For every solution $T_{\rm C}$ of \eqref{eq:Coulomb5th}, in particular for the unrestricted four-parameter family,
\begin{equation}
\comm{H_{\rm C}}{Y_{\rm C}}=0.
\label{eq:HCYCzero}
\end{equation}
\end{theorem}
For an arbitrary function $T_{\rm C}$, normal ordering gives $\comm{H_{\rm C}}{Y_{\rm C}}=\tfrac12\hb^2r^{-3}\acom{\mathcal E_5^{\rm C}[T_{\rm C}]}{\partial_\theta}$, where $\mathcal E_5^{\rm C}[T_{\rm C}]$ is the right-hand side of \eqref{eq:Coulomb5th}.  Hence \eqref{eq:HCYCzero} holds if and only if \eqref{eq:Coulomb5th} does.

In normal order, at $\hb=1$ and with $s=\sin\theta$, $c=\cos\theta$, $\tau_j=T_{\rm C}^{(j)}$, \eqref{eq:CYcorrected} reads
\begin{align}
Y_{\rm C}={}&2s\,\partial_r\partial_\theta^3+\frac{2c}{r}\partial_\theta^4+3c\,\partial_r\partial_\theta^2-\frac{3s}{r}\partial_\theta^3
\nonumber\\
&-\bigl[(6\tau_1+4c_{32}+3)s+(2\tau_0+4c_{31})c\bigr]\partial_r\partial_\theta
+\frac1r\bigl[(2\tau_0+4c_{31})s-(8\tau_1+4c_{32}+3)c-2acr\bigr]\partial_\theta^2\nonumber\\
&+\bigl[(\tau_0+2c_{31}-3\tau_2)s-(4\tau_1+2c_{32}+1)c\bigr]\partial_r\nonumber\\
&+\frac1r\bigl[(7\tau_1+2c_{32}+1+2ar)s+(\tau_0+2c_{31}-8\tau_2)c\bigr]\partial_\theta
+a\bigl[(4\tau_1+4c_{32}+2)c-(2\tau_0+4c_{31})s\bigr]\nonumber\\
&+\frac1{2r}\bigl[(\tau_4+3\tau_2-12\tau_1\tau_2-8c_{32}\tau_2)s-(4\tau_3-2\tau_1+4\tau_0\tau_2+8c_{31}\tau_2)c\bigr].
\label{eq:CYnormal}
\end{align}

Commuting this operator with $H_{\rm C}$, normal ordering the result and eliminating the higher derivatives of $T_{\rm C}$ with \eqref{eq:Cjet3} and \eqref{eq:Cderivation} gives
\begin{equation}
\comm{H_{\rm C}}{X_{\rm C}}=0,\qquad
\comm{H_{\rm C}}{Y_{\rm C}}=0.
\end{equation}

\paragraph{Lower-order terms from the polar transformation.}

The identity \eqref{eq:polarorderingidentity} requires no Painlev\'e equation.  Starting from
\begin{equation}
\partial_x=\cos\theta\,\partial_r-\frac{\sin\theta}{r}\partial_\theta,
\qquad
\partial_y=\sin\theta\,\partial_r+\frac{\cos\theta}{r}\partial_\theta,
\end{equation}
one inserts these expressions into the Cartesian symmetrized quadratic block and normal orders all $r$ and $\theta$ derivatives.  Differentiation of the coordinate-dependent coefficients generates the last line of \eqref{eq:polarorderingidentity}.  For $G_1=0$ the additional first-order term is $G_3\partial_\theta/r$, accompanied by the scalar correction in \eqref{eq:CYcorrected}.

The ordinary Coulomb case $T_{\rm C}=c_{31}=c_{32}=0$ shows what these terms do.  There the operator reduces to $Y_{\rm C}=\acom{L^2}{A_x}+\hb^2A_x$, where $L=-i\hb\partial_\theta$ and $A_x=\tfrac12\acom{p_y}{L}+ax/r$ is the quantum Runge--Lenz component, and this fixes the sign of the terminal term $-a\hb^2\cos\theta$ in \eqref{eq:CG4source}.  With the opposite sign, which is what one obtains by dropping the lower-order polar terms, $\comm{H_{\rm C}}{Y_{\rm C}}$ has the nonzero remainder $2a\hb^2\cos\theta$.

\section{Coulomb branch: polynomial algebra and Heun--Racah structure}
\label{sec:calg}
The nested commutators of the Coulomb integral give the second cubic algebra.  Set
\begin{equation}
 A_{\rm C}=X_{\rm C}-\frac{\hb^2}{4},\qquad B_{\rm C}=Y_{\rm C},\qquad
 C_{\rm C}=\comm{A_{\rm C}}{B_{\rm C}}.
 \label{eq:CABC}
\end{equation}
The first nested commutator fixes the quarter-shift.  The algebra is
\begin{align}
\comm{A_{\rm C}}{B_{\rm C}}={}&C_{\rm C},
\label{eq:Calg1}\\
\comm{A_{\rm C}}{C_{\rm C}}={}&2\hb^2\acom{A_{\rm C}}{B_{\rm C}}
-8\hb^6a q_8^{\rm C},
\label{eq:Calg2}\\
\comm{B_{\rm C}}{C_{\rm C}}={}&-2\hb^2B_{\rm C}^2+64\hb^2H_{\rm C}A_{\rm C}^3\nonumber\\
&+\left[32\hb^4(6q_7^{\rm C}+5)H_{\rm C}+24\hb^2a^2\right]A_{\rm C}^2\nonumber\\
&+\left[32\hb^6(8q_7^{\rm C}+3-2q_8^{\rm C}-4q_9^{\rm C})H_{\rm C}
+32\hb^4a^2(2q_7^{\rm C}+1)\right]A_{\rm C}\nonumber\\
&+16\hb^8(\Delta_{\rm C}+4q_{10}^{\rm C})H_{\rm C}
+8\hb^6a^2\Delta_{\rm C}.
\label{eq:Calg3}
\end{align}
Because $q_8^{\rm C}=-K_1/\hb^4$, the central term in \eqref{eq:Calg2} is equivalently $8\hb^2aK_1$.

\begin{theorem}[Coulomb $P_{\rm VI}$ algebra]
Equations \eqref{eq:Calg1}--\eqref{eq:Calg3} hold for every solution of the four-parameter Coulomb SD-I.a/$P_{\rm VI}$ angular equation.
\end{theorem}

At fixed $H_{\rm C}=E$, \eqref{eq:Calg1}--\eqref{eq:Calg3} are again a Heun--Racah specialization.  In the notation of \eqref{eq:HR1}--\eqref{eq:HR2},
\begin{equation}
g_2=2\hb^2,\qquad g_5=-8\hb^6a q_8^{\rm C},\qquad
 e_2=-64\hb^2E,
\label{eq:CHeunKey}
\end{equation}
with $e_1,g_6,g_7$ obtained by changing the signs of the coefficients of $A_{\rm C}^2,A_{\rm C},1$ in \eqref{eq:Calg3}.  The oscillator and Coulomb algebras therefore lie in the same Heun--Racah family, but the cubic Heun coefficient is proportional to $\lambda$ for the oscillator and to the fixed energy $E$ for the Coulomb system.

\paragraph{Derivation of the relations.}
The reduction of Section~\ref{sec:cop} proves the two polynomial-algebra relations: after normal ordering and elimination of the higher derivatives of $T_{\rm C}$ through \eqref{eq:Cjet3} and \eqref{eq:Cderivation}, every differential coefficient in \eqref{eq:Calg2} and \eqref{eq:Calg3} vanishes.  The last lower-order terms can also be understood without expanding them.  After cancellation through order two, the remaining residual has order at most one and, by Jacobi's identity, commutes with both $A_{\rm C}$ and $H_{\rm C}$.  Write it as
\begin{equation}
R=P(r,\theta)\partial_r+Q(r,\theta)\partial_\theta+S(r,\theta).
\end{equation}
The highest angular derivatives in $\comm{A_{\rm C}}{R}=0$ force $P_\theta=Q_\theta=0$, and the coefficient of $T_{\rm C}''$ forces $Q=0$.  Then $\comm{H_{\rm C}}{R}=0$ forces $P$ to be constant and the derivative of $a/r+T_{\rm C}'/r^2$ forces $P=0$.  The remaining function $S$ satisfies $\partial_rS=\mathscr D_{\rm C}S=0$, so it is a first integral of the angular flow rather than necessarily a number.  The zero-order coefficient of \eqref{eq:Calg3} then gives $S=0$.  Therefore $R=0$.

\paragraph{Coulomb Casimir.}
Both \eqref{eq:alg1}--\eqref{eq:alg3} and \eqref{eq:Calg1}--\eqref{eq:Calg3} belong to the general class
\begin{equation}
[A,B]=C,\qquad [A,C]=\mathfrak b\acom{A}{B}+\mathfrak f,\qquad
[B,C]=-\mathfrak bB^2+\mathfrak gA^3+\mathfrak hA^2+\mathfrak iA+\mathfrak j,
\end{equation}
with central coefficients.  Ordering the products with the defining relations gives the central element
\begin{align}
\mathcal K={}&C^2-\mathfrak b\acom{A}{B^2}+\mathfrak b^2B^2-2\mathfrak fB+\frac{\mathfrak g}2A^4
+\frac{2(\mathfrak b\mathfrak g+\mathfrak h)}3A^3\nonumber\\
&-\frac{\mathfrak b^2\mathfrak g-2\mathfrak b\mathfrak h-6\mathfrak i}{6}A^2+2\mathfrak jA,
\label{eq:GeneralCasimir}
\end{align}
which reproduces \eqref{eq:Casimir} for the oscillator coefficients \eqref{eq:coeff1}--\eqref{eq:coeff5}.  For the Coulomb coefficients it becomes
\begin{align}
\mathcal K_{\rm C}={}&C_{\rm C}^2-2\hb^2\acom{A_{\rm C}}{B_{\rm C}^2}
+4\hb^4B_{\rm C}^2+16\hb^6a q_8^{\rm C}B_{\rm C}
+32\hb^2H_{\rm C}A_{\rm C}^4\nonumber\\
&+\left[64\hb^4(2q_7^{\rm C}+3)H_{\rm C}+16\hb^2a^2\right]A_{\rm C}^3\nonumber\\
&+\left[32\hb^6(12q_7^{\rm C}+5-2q_8^{\rm C}-4q_9^{\rm C})H_{\rm C}
+16\hb^4a^2(4q_7^{\rm C}+3)\right]A_{\rm C}^2\nonumber\\
&+\left[32\hb^8(\Delta_{\rm C}+4q_{10}^{\rm C})H_{\rm C}
+16\hb^6a^2\Delta_{\rm C}\right]A_{\rm C}.
\label{eq:CCasimir}
\end{align}
Writing all products in ordered form gives $[A_{\rm C},\mathcal K_{\rm C}]=[B_{\rm C},\mathcal K_{\rm C}]=0$.  Substituting the Coulomb differential operators and reducing every coefficient from differential order ten through zero gives
\begin{equation}
\mathcal K_{\rm C}=-32\hb^{10}(q_8^{\rm C})^2H_{\rm C}
-64\hb^8a^2q_{10}^{\rm C}.
\label{eq:CCasimirValue}
\end{equation}

\section{Coulomb branch: representations and spectrum}
\label{sec:cspec}
At fixed $H_{\rm C}=E$ use the same deformed-oscillator relations \eqref{eq:oscdef} and write $t=N+u$.  The first Coulomb adjoint relation gives
\begin{equation}
A_{\rm C}(t)=\hb^2\left(t^2-\frac14\right),\qquad
X_{\rm C}(t)=\hb^2t^2,
\label{eq:CAofN}
\end{equation}
and its diagonal part gives
\begin{equation}
d_{\rm C}(t)=\frac{2\hb^2a q_8^{\rm C}}{t^2-1/4}.
\label{eq:CdofN}
\end{equation}
Choose the same convenient gauge
\begin{equation}
\rho_{\rm C}(t)^2=\frac1{t(t+1)(2t+1)^2},\qquad \xi=2t-1.
\end{equation}
Solving the diagonal part of \eqref{eq:Calg3} together with the Casimir value \eqref{eq:CCasimirValue} gives the gauge-invariant transition product
\begin{equation}
\mathcal A_{\rm C}(t)=\rho_{\rm C}(t-1)^2\Phi_{\rm C}(t)
=\frac{\hb^4\left(2a^2+\hb^2E\xi^2\right)R_4(\xi^2)}
{128\,t(t-1)\xi^2}.
\label{eq:Ctransition}
\end{equation}
Hence
\begin{equation}
\Phi_{\rm C}(t)=\frac{\hb^4}{128}
\left(2a^2+\hb^2E\xi^2\right)R_4(\xi^2).
\label{eq:CPhi}
\end{equation}
The angular quartic is the same $R_4$ as in the oscillator structure function \eqref{eq:Phi}, with $q_k$ replaced by $q_k^{\rm C}$. Then the two systems differ only in the radial factor,
\begin{equation}
E^2-2\lambda\hb^2\xi^2\quad\text{versus}\quad
2a^2+\hb^2E\xi^2.
\label{eq:radialfactorcomparison}
\end{equation}

\paragraph{Attractive modules and radial separation.}
For $E<0$ define
\begin{equation}
 r_{\rm C}=\frac{\sqrt2\,|a|}{\hb\sqrt{-E}}.
\label{eq:rC}
\end{equation}
Then the radial factor in \eqref{eq:CPhi} is proportional to $r_{\rm C}^2-\xi^2$.  If the four real angular moduli $a_j$ are those of \eqref{eq:amoduli}, the Coulomb transition roots are
\begin{equation}
\mathscr R_{\rm C}=\{\pm r_{\rm C},\ \pm a_1,\ldots,\pm a_4\}.
\label{eq:Crootset}
\end{equation}
A finite-dimensional module is obtained by choosing two roots separated by $2P$, $P\in\mathbb N$, and requiring $(r_{\rm C}^2-\xi_n^2)\prod_{j=1}^4(\xi_n^2-a_j^2)/(\xi_n^2-1)>0$ at every interior lattice point $\xi_n=\xi_0+2n$, for lattices avoiding $\{-3,-2,-1,0,1\}$.  This is the Coulomb counterpart of the oscillator criterion in Section~\ref{sec:energy}.

For the attractive problem, the branch starting at the largest positive angular modulus $a_{\max}$ terminates at $+r_{\rm C}$.  Put $P=p+1$.  The endpoint condition $r_{\rm C}=a_{\max}+2P$ gives
\begin{equation}
E_p=-\frac{2a^2}{\hb^2(a_{\max}+2p+2)^2},\qquad \dim\mathcal H_p=p+1.
\label{eq:Cexpectedenergy}
\end{equation}
For every interior point of this branch, $a_j^2<\xi_n^2<r_{\rm C}^2$, so the radial factor and all four angular factors are positive, as is $\xi_n^2-1$ because $\xi_n\ge2$.  The module is therefore algebraically unitary for real monodromy data away from the exceptional loci.  As for the oscillator, it gives the physical shell when the angular ground state corresponds to $a_{\max}$.

The same energy formula follows from radial separation.  If $X_{\rm C}$ has eigenvalue $x$, then
\begin{equation}
 \left[-\frac{\hb^2}{2}\left(\frac{\dd^2}{\dd r^2}+\frac1r\frac{\dd}{\dd r}\right)
 +\frac{x}{2r^2}+\frac{a}{r}\right]R(r)=ER(r),
 \label{eq:Cradial}
\end{equation}
and regular attractive bound states satisfy
\begin{equation}
 E=-\frac{a^2}{2\hb^2\left(n_r+\frac12+\frac{\sqrt{x}}{\hb}\right)^2}.
 \label{eq:Cenergyradial}
\end{equation}
From \eqref{eq:CAofN}, $\sqrt{x_n}/\hb=n+u$.  With $u=(a_{\max}+1)/2$ and fixed $p=n+n_r$, equation \eqref{eq:Cenergyradial} reproduces \eqref{eq:Cexpectedenergy}.

\paragraph{Angular and radial domains.}
The separated oscillator and Coulomb radial problems share the same inverse-square behavior at the origin.  If $X$ has eigenvalue $x$, set $R(r)=r^{-1/2}u(r)$.  The radial operator on $L^2(\mathbb R_+,\dd r)$ becomes
\begin{equation}
-\frac{\hb^2}{2}u''(r)+\frac{x-\hb^2/4}{2r^2}u(r)+V_{\rm rad}(r)u(r),
\label{eq:radialhalfline}
\end{equation}
with $V_{\rm rad}(r)=\lambda r^2$ or $a/r$.  Hence $x\ge0$ is the natural semibounded regime for the Friedrichs realization.  For $x\ge\hb^2$ the inverse-square endpoint is limit point.  For $0\le x<\hb^2$ both local behaviors are square integrable and a self-adjoint boundary condition must be chosen.  The energy formulas above correspond to the regular/Friedrichs branch \cite{DerezinskiRichard}.

The angular operator $X=-\hb^2\partial_\theta^2+2T'(\theta)$ is a separate one-dimensional spectral problem.  For a real pole-free $P_{\rm VI}$ branch on a compact interval, standard regular boundary conditions give a self-adjoint realization.  If the chosen branch has inverse-square singular endpoints, a local limit-point/limit-circle analysis is required.  Positivity of the deformed-oscillator structure function is an algebraic unitarity condition. However, the physical angular domain must still be specified separately.

\section{Classical comparison}
\label{sec:classical}
We now compare the quantum results with the classical Poisson algebras of \cite{ELW2017}.  We keep the source convention $C^{\rm cl}=\pb{Y^{\rm cl}}{X^{\rm cl}}$, whose orientation is opposite to the quantum convention $C=[A,B]$.  Since the classical angular equations arise from a singular $\hb\to0$ limit, the Poisson algebras are included for comparison and are not used to obtain the quantum coefficients.

\paragraph{Oscillator Poisson algebra.}
In the oscillator branch the classical limit of the angular equation is singular: after integration, the fifth-order quantum equation reduces to a first-order nonlinear equation that does not have the Painlev\'e property \cite{ELW2017}, so there is no classical $P_{\rm VI}$ family corresponding term by term to the quantum one.

For the oscillator-type classical branch write, with $b$ the classical counterpart of $\lambda$,
\begin{align}
H_{\rm O}^{\rm cl}&=\frac12\left(p_r^2+\frac{p_\theta^2}{r^2}\right)+br^2+\frac{(T_{\rm O}^{\rm cl})'(\theta)}{r^2},\\
X_{\rm O}^{\rm cl}&=p_\theta^2+2(T_{\rm O}^{\rm cl})'(\theta).
\end{align}
With $z=\tan 2\theta$, the primitive $T=T_{\rm O}^{\rm cl}$ satisfies the first-order equation
\begin{align}
0={}&T^2-(c_{11}+2zT_z)T-3z^2(z^2+1)T_z^2+z(c_{12}z+c_{11})T_z
-\frac{K_1}{4\sqrt{z^2+1}}+K_2.
\label{eq:classOscODE}
\end{align}
The constants here are classical: with $\hb$ restored in \eqref{eq:jetflow} as in \eqref{eq:qmaphbar}, multiplying the difference of the two sides of \eqref{eq:jetflow} by $\hb^4$ and letting $\hb\to0$ gives $8$ times the right-hand side of \eqref{eq:classOscODE}, with $c_{11}^{\rm cl}=c_{11}/2$, $c_{12}^{\rm cl}=c_{12}/2$, $K_1^{\rm cl}=K_1/2$ and $K_2^{\rm cl}=K_2$ in terms of the quantum constants, where $\cos2\theta>0$ so that $1/\sqrt{z^2+1}=\cos2\theta$.  For $\cos2\theta<0$, one has instead $K_1^{\rm cl}=-K_1/2$.  The fourth-order integral may be written
\begin{align}
Y_{\rm O}^{\rm cl}={}&2p_r^2p_\theta^2\cos2\theta-\frac{2}{r^2}p_\theta^4\cos2\theta
-\frac4r p_rp_\theta^3\sin2\theta\nonumber\\
&+2G_1p_r^2+2G_3p_rp_\theta+2G_2p_\theta^2+G_4,
\end{align}
where
\begin{align}
G_1={}&2\cos2\theta\,T'-2\sin2\theta\,T+c_{11}\sin2\theta-c_{12}\cos2\theta,\\
G_2={}&\frac{2\sin2\theta\,T-4\cos2\theta\,T'-c_{11}\sin2\theta+c_{12}\cos2\theta}{r^2}+2br^2\cos2\theta,\\
G_3={}&\frac{2c_{11}\cos2\theta-4\cos2\theta\,T-6\sin2\theta\,T'+2c_{12}\sin2\theta}{r},
\end{align}
and
\begin{align}
G_4={}&\frac1{r^2}\Bigl(\bigl[(4T-2c_{11})\cos2\theta+(6T'-2c_{12})\sin2\theta\bigr]T''\nonumber\\
&\qquad +(8T'-4c_{12})T'\cos2\theta-(8T-4c_{11})T'\sin2\theta\Bigr)\nonumber\\
&+4br^2\bigl(2\cos2\theta\,T'-2\sin2\theta\,T+c_{11}\sin2\theta-c_{12}\cos2\theta\bigr).
\end{align}
Define $C_{\rm O}^{\rm cl}=\pb{Y_{\rm O}^{\rm cl}}{X_{\rm O}^{\rm cl}}$.  On the locus
\begin{equation}
K_1=c_{12}=0,\qquad K_2=\frac14c_{11}^2,
\end{equation}
equation \eqref{eq:classOscODE} becomes $(T-\tfrac12c_{11})^2-2zT_z(T-\tfrac12c_{11})-3z^2(z^2+1)T_z^2=0$, which is homogeneous in $T-\tfrac12c_{11}$ and has the two explicit solutions
\begin{equation}
T=\frac{c_{11}}2+\zeta\left[\frac{z(w+1)}{(w+2)^2}\right]^{1/3},\qquad
T=\frac{c_{11}}2+\zeta\left[\frac{(w-1)(w+2)}{z(w-2)}\right]^{1/3},\qquad
w=\sqrt{3z^2+4},
\label{eq:classOscFactorized}
\end{equation}
with $\zeta$ a free constant.  For these two solutions the algebra reduces to
\begin{align}
\pb{X_{\rm O}^{\rm cl}}{C_{\rm O}^{\rm cl}}&=16X_{\rm O}^{\rm cl}Y_{\rm O}^{\rm cl},\\
\pb{Y_{\rm O}^{\rm cl}}{C_{\rm O}^{\rm cl}}&=8\left[48(H_{\rm O}^{\rm cl})^2(X_{\rm O}^{\rm cl})^2
-128b(X_{\rm O}^{\rm cl})^3-(Y_{\rm O}^{\rm cl})^2+b\sigma_{3,4}\right],
\label{eq:classOscSpecial}
\end{align}
with, respectively,
\begin{equation}
\sigma_3=\frac{512}{9}\zeta^3,\qquad \sigma_4=-\frac{512}{3}\zeta^3.
\end{equation}
For every solution of \eqref{eq:classOscODE} the full closure is
\begin{align}
\pb{X_{\rm O}^{\rm cl}}{C_{\rm O}^{\rm cl}}
={}&16X_{\rm O}^{\rm cl}Y_{\rm O}^{\rm cl}-32K_1H_{\rm O}^{\rm cl},
\label{eq:classOsc1}\\
\pb{Y_{\rm O}^{\rm cl}}{C_{\rm O}^{\rm cl}}
={}&8\Bigl[48(H_{\rm O}^{\rm cl})^2(X_{\rm O}^{\rm cl})^2-(Y_{\rm O}^{\rm cl})^2-128b(X_{\rm O}^{\rm cl})^3\nonumber\\
&\quad-64c_{12}(H_{\rm O}^{\rm cl})^2X_{\rm O}^{\rm cl}
+16(c_{11}^2+c_{12}^2-4K_2)(H_{\rm O}^{\rm cl})^2
+192bc_{12}(X_{\rm O}^{\rm cl})^2\Bigr]\nonumber\\
&-512b(c_{11}^2+c_{12}^2-4K_2)X_{\rm O}^{\rm cl}-b\Lambda_{\rm O}.
\label{eq:classOsc2}
\end{align}
Here $\Lambda_{\rm O}$ is the value, on the chosen solution, of a first integral of \eqref{eq:classOscODE}.  The special algebra \eqref{eq:classOscSpecial} is recovered from \eqref{eq:classOsc1}--\eqref{eq:classOsc2} by the special parameter values above together with $\Lambda_{\rm O}=-8\sigma_{3,4}$.  The products $XY$, $Y^2$ and $X^3$ show the same degree pattern as the quantum algebra, although the two sets of coefficients come from differential problems related only through the singular limit.

\paragraph{Coulomb Poisson algebra.}
For the Coulomb branch, set
\begin{align}
H_{\rm C}^{\rm cl}&=\frac12\left(p_r^2+\frac{p_\theta^2}{r^2}\right)+\frac{a}{r}+\frac{(T_{\rm C}^{\rm cl})'(\theta)}{r^2},\\
X_{\rm C}^{\rm cl}&=p_\theta^2+2(T_{\rm C}^{\rm cl})'(\theta).
\end{align}
With $z=\tan\theta$, $T=T_{\rm C}^{\rm cl}$ obeys
\begin{align}
0={}&T^2-2(zT_z-2c_{31})T-3z^2(z^2+1)T_z^2-4z(c_{32}z+c_{31})T_z
+\frac{2K_1}{\sqrt{z^2+1}}-\frac{K_2}{2}.
\label{eq:classCoulODE}
\end{align}
Here the constants coincide with those of \eqref{eq:Cjet3} where $\cos\theta>0$, so that $1/\sqrt{z^2+1}=\cos\theta$.  For $\cos\theta<0$, the sign of $K_1$ is reversed.  Multiplying the difference of the two sides of \eqref{eq:Cjet3} by $\hb^4\sin^2\theta$ and letting $\hb\to0$ gives $2$ times the right-hand side of \eqref{eq:classCoulODE}.  The fourth-order classical integral is
\begin{align}
Y_{\rm C}^{\rm cl}={}&2\sin\theta\,p_\theta^3p_r+\frac{2\cos\theta}{r}p_\theta^4
+2G_1p_r^2+2G_3p_rp_\theta+2G_2p_\theta^2+G_4,
\end{align}
with
\begin{align}
G_1={}&0,\\
G_2={}&\frac{4\cos\theta\,T'+2c_{32}\cos\theta-(T+2c_{31})\sin\theta}{r}+a\cos\theta,\\
G_3={}&3\sin\theta\,T'+(T+2c_{31})\cos\theta+2c_{32}\sin\theta,\\
G_4={}&-\frac2r\bigl(3\sin\theta\,T'+\cos\theta\,T+2c_{31}\cos\theta+2c_{32}\sin\theta\bigr)T''\nonumber\\
&+2a(2\cos\theta\,T'-\sin\theta\,T)-4a(c_{31}\sin\theta-c_{32}\cos\theta).
\end{align}
Define
\begin{equation}
C_{\rm C}^{\rm cl}=\pb{Y_{\rm C}^{\rm cl}}{X_{\rm C}^{\rm cl}}.
\end{equation}
On $K_1=c_{32}=0$, $K_2=-8c_{31}^2$, equation \eqref{eq:classCoulODE} becomes $(T+2c_{31})^2-2zT_z(T+2c_{31})-3z^2(z^2+1)T_z^2=0$, with the two explicit solutions $T=-2c_{31}+\zeta[\,\cdot\,]^{1/3}$ given by the same two radicals as in \eqref{eq:classOscFactorized}, now with $z=\tan\theta$.  For these,
\begin{align}
\pb{X_{\rm C}^{\rm cl}}{C_{\rm C}^{\rm cl}}&=4X_{\rm C}^{\rm cl}Y_{\rm C}^{\rm cl},\\
\pb{Y_{\rm C}^{\rm cl}}{C_{\rm C}^{\rm cl}}&=2\left[32H_{\rm C}^{\rm cl}(X_{\rm C}^{\rm cl})^3
+12a^2(X_{\rm C}^{\rm cl})^2-(Y_{\rm C}^{\rm cl})^2+\sigma_{1,2}H_{\rm C}^{\rm cl}\right],
\label{eq:classCoulSpecial}
\end{align}
where, respectively, $\sigma_1=-16\zeta^3/9$ and $\sigma_2=16\zeta^3/3$.  For every solution of \eqref{eq:classCoulODE},
\begin{align}
\pb{X_{\rm C}^{\rm cl}}{C_{\rm C}^{\rm cl}}
={}&4X_{\rm C}^{\rm cl}Y_{\rm C}^{\rm cl}+8aK_1,
\label{eq:Cclass1}\\
\pb{Y_{\rm C}^{\rm cl}}{C_{\rm C}^{\rm cl}}
={}&2\Bigl[32H_{\rm C}^{\rm cl}(X_{\rm C}^{\rm cl})^3-(Y_{\rm C}^{\rm cl})^2
+12a^2(X_{\rm C}^{\rm cl})^2+96c_{32}H_{\rm C}^{\rm cl}(X_{\rm C}^{\rm cl})^2\nonumber\\
&\qquad+64\left(c_{31}^2+c_{32}^2+\frac{K_2}{8}\right)H_{\rm C}^{\rm cl}X_{\rm C}^{\rm cl}
+32a^2c_{32}X_{\rm C}^{\rm cl}\Bigr]\nonumber\\
&-\Lambda_{\rm C}H_{\rm C}^{\rm cl}
+32a^2\left(c_{31}^2+c_{32}^2+\frac{K_2}{8}\right).
\label{eq:Cclass2}
\end{align}
Here $\Lambda_{\rm C}$ is the value, on the chosen solution, of a first integral of \eqref{eq:classCoulODE}.  The special algebra \eqref{eq:classCoulSpecial} has $\Lambda_{\rm C}=-2\sigma_{1,2}$.

\section{Comparison of the two quantum systems}
\label{sec:unified}
The main similarities and differences between the two branches are summarized below.
\begin{center}
\small
\begin{tabular}{>{\raggedright\arraybackslash}p{3.0cm}>{\raggedright\arraybackslash}p{5.3cm}>{\raggedright\arraybackslash}p{5.3cm}}
\toprule
feature & oscillator $B_3$ branch & extended Coulomb $A_2$ branch\\
\midrule
quantum radial term & $\lambda r^2$ & $a/r$\\
quantum angular equation & fifth-order, four-parameter SD-I.a/$P_{\rm VI}$ & fifth-order, four-parameter SD-I.a/$P_{\rm VI}$\\
quantum fourth-order integral & explicit normal-ordered oscillator integral & Cartesian Hermitian integral transformed to its full polar operator form\\
quantum algebra & Heun--Racah cubic closure \eqref{eq:alg1}--\eqref{eq:alg3} & Heun--Racah cubic closure \eqref{eq:Calg1}--\eqref{eq:Calg3}\\
quantum Casimir & \eqref{eq:Casimir}, value \eqref{eq:CasimirValue} & \eqref{eq:CCasimir}, value \eqref{eq:CCasimirValue}\\
angular transition factor & $R_4(\xi^2)$ & $R_4(\xi^2)$ with $q_k$ replaced by $q_k^{\rm C}$\\
radial transition factor & $E^2-2\lambda\hb^2\xi^2$ & $2a^2+\hb^2E\xi^2$\\
classical angular equation & formal $\hb\to0$ limit of \eqref{eq:jetflow}, with singular SD-I.a constants & formal $\hb\to0$ limit of \eqref{eq:Cjet3}, with singular SD-I.a constants\\
\bottomrule
\end{tabular}
\end{center}

Both quantum algebras are central specializations of the Heun--Racah algebra.  The oscillator cubic coefficient is fixed by $\lambda$, whereas the Coulomb coefficient is proportional to the central energy.  Their structure functions have the same angular quartic $R_4(\xi^2)$ but different radial factors, as displayed in the table.  The Levi-Civita correspondence makes this relation explicit.

\paragraph{Levi-Civita correspondence.}
Write $(r',\theta')=(r^2,2\theta)$ for polar coordinates in the Coulomb plane.  With $H_{\rm C}$ written in $(r',\theta')$,
\[
4r'(H_{\rm C}-E)=H+4a,
\]
where $H$ is the oscillator Hamiltonian \eqref{eq:H} with $\lambda=-4E$ and $T(\theta)=2T_{\rm C}(2\theta)$.  This is the oscillator--Coulomb coupling-constant metamorphosis \cite{HGDR1984} in the form given by Post and Winternitz \cite{PostWinternitz2010}, valid for an arbitrary angular potential. The general theory for such transformations is in \cite{KalninsMillerPost2010}.

For the present $P_{\rm VI}$ systems the map is more explicit.  Equation \eqref{eq:PVI5th} becomes $-32$ times \eqref{eq:Coulomb5th} under
\[
c_{11}=-16c_{31},\qquad c_{12}=-4\hb^2-16c_{32},
\]
and the constants $(K_1,K_2)$ of \eqref{eq:jetflow} map to $(-64K_1^{\rm C},-2K_2^{\rm C})$.  The SD-I.a parameters in \eqref{eq:qmaphbar} then become those in \eqref{eq:Cqmap}.

The algebra follows the same correspondence.  Under
\begin{center}
\begin{tabular}{cc}
\toprule
oscillator datum & Coulomb image\\
\midrule
$A$ & $4A_{\rm C}$\\
$B$ & $-32B_{\rm C}$\\
$H$ & $-4a$\\
$\lambda$ & $-4H_{\rm C}$\\
\bottomrule
\end{tabular}
\end{center}
the oscillator relations become the Coulomb relations, the Casimir values differ by the factor $16384$, and the transition product in \eqref{eq:transition} becomes $1024$ times \eqref{eq:Ctransition}.  At the operator level, replacing $\lambda$ in $Y$ by $-4H_{\rm C}$ on the right of $\mathcal U_\lambda$, as prescribed by Theorem~2 of \cite{PostWinternitz2010}, gives $-32Y_{\rm C}$.  This correspondence explains both the shift $A_{\rm C}=X_{\rm C}-\hb^2/4$ and the common quartic factor $R_4$.  The physical interpretations of the two algebras nevertheless differ, because a coupling on one side becomes the energy on the other.

\section{Conclusion}
We have derived the quantum polynomial algebras associated with the two fourth-order polar systems whose angular potentials are governed by the general four-parameter Painlev\'e~VI transcendent.  In the oscillator branch the second- and fourth-order symmetries generate a cubic Heun--Racah algebra with the Casimir \eqref{eq:CasimirValue} and a deformed-oscillator representation.  When the angular ground state corresponds to the largest angular root $a_{\max}$, the positive-energy module gives
\[
E=\hbar\sqrt{2\lambda}\,[a_{\max}+2(p+1)],
\]
in agreement with radial separation.

For the Coulomb branch, the Cartesian Hermitian integral must first be transformed to its full polar operator form.  The resulting fourth-order symmetry generates the cubic algebra \eqref{eq:Calg1}--\eqref{eq:Calg3}, with Casimir value \eqref{eq:CCasimirValue} and deformed-oscillator structure function \eqref{eq:CPhi}.  Under the same condition on the angular ground state, the attractive module gives
\[
E_p=-\frac{2a^2}{\hbar^2(a_{\max}+2p+2)^2},
\]
again agreeing with radial separation.

The two structure functions take the parallel forms
\begin{align*}
\Phi_{\rm O}&\propto\left(E^2-2\lambda\hb^2\xi^2\right)R_4(\xi^2),\\
\Phi_{\rm C}&\propto\left(2a^2+\hb^2E\xi^2\right)R_4(\xi^2).
\end{align*}
The common quartic $R_4$ contains the Painlev\'e~VI angular data, while the remaining factor records the radial dynamics.  The Levi-Civita correspondence explains this parallel structure and maps the oscillator algebra, Casimir and representation directly to their Coulomb counterparts.  These results show that the finite quantum symmetry algebra is not restricted to rational Painlev\'e potentials.

At the algebraic level the representation theory is settled by Theorem~\ref{thm:lattice} and its Coulomb counterpart: for generic parameters, every irreducible unitary module at fixed energy is obtained by choosing two transition roots with spacing $2P$ and checking a finite set of sign conditions, and the Painlev\'e~VI data enter only through the roots $\pm(\theta_0\pm\theta_1)$, $\pm(\theta_\infty\pm\theta_x)$ of $R_4$.  Which of these modules occur in the Schr\"odinger spectrum is a question about the angular operator $X$ alone, through the real Painlev\'e~VI branch, the interval between its poles and the self-adjoint realization chosen there.  Two questions remain open even within the present family: which real branches and boundary conditions select each algebraic angular root, and how the resonant modules at special monodromy values are realized spectrally.  On the classical side, the Poisson algebras of \cite{ELW2017} have the degree pattern of the quantum relations, but since the classical angular equation loses the Painlev\'e property they are not term-by-term limits of them.

Two further directions follow from the structure found here.  The third-order polar system of \cite{TremblayWinternitz2010}, whose angular potential is also a Painlev\'e~VI transcendent, and the higher-order polar families of \cite{ERWY2018} are natural places to ask whether the symmetry algebra is again a central specialization of Heun--Racah type, with the radial dynamics confined to one factor of the structure function.  Separately, in \cite{VinetZhedanovHeunHahn,BergeronEtAl} the Heun--Racah algebra is realized by algebraic Heun operators attached to the Hahn and Racah bispectral problems. It would be interesting to investigate whether the angular Painlev\'e~VI operator admits a comparable tridiagonal realization, at least on the rational loci of \cite{MPR2020,MPR2022}.

\section*{Data availability}
No new data were created or analysed in this study. Maple and Mathematica scripts that reproduce the operator identities, polynomial algebras, Casimir values and deformed-oscillator formulas of this paper are available upon request from the authors.

\section*{Acknowledgements}
IM was supported by the Australian Research Council Future Fellowship FT180100099.


\begin{thebibliography}{99}
\bibitem{MillerPostWinternitz}
W.~Miller Jr, S.~Post and P.~Winternitz,
Classical and quantum superintegrability with applications,
\emph{J. Phys. A: Math. Theor.} \textbf{46} 423001 (2013).

\bibitem{BonatsosDaskaloyannisKokkotas}
D.~Bonatsos, C.~Daskaloyannis and K.~Kokkotas,
Deformed oscillator algebras for two-dimensional quantum superintegrable systems,
\emph{Phys. Rev. A} \textbf{50} 3700--3709 (1994).

\bibitem{Daskaloyannis}
C.~Daskaloyannis,
Quadratic Poisson algebras of two-dimensional classical superintegrable systems and quadratic associative algebras of quantum superintegrable systems,
\emph{J. Math. Phys.} \textbf{42} 1100--1119 (2001).

\bibitem{MarquetteQuartic}
I.~Marquette,
Quartic Poisson algebras and quartic associative algebras and realizations as deformed oscillator algebras,
\emph{J. Math. Phys.} \textbf{54} 071702 (2013).

\bibitem{GravelWinternitz2002}
S.~Gravel and P.~Winternitz,
Superintegrability with third-order integrals in quantum and classical mechanics,
\emph{J. Math. Phys.} \textbf{43} 5902--5912 (2002).

\bibitem{Gravel2004}
S.~Gravel,
Hamiltonians separable in Cartesian coordinates and third-order integrals of motion,
\emph{J. Math. Phys.} \textbf{45} 1003--1019 (2004).

\bibitem{MarquetteWinternitz2007}
I.~Marquette and P.~Winternitz,
Polynomial Poisson algebras for classical superintegrable systems with a third-order integral of motion,
\emph{J. Math. Phys.} \textbf{48} 012902 (2007).

\bibitem{Marquette2009}
I.~Marquette,
Superintegrability with third order integrals of motion, cubic algebras, and supersymmetric quantum mechanics. II. Painlev\'e transcendent potentials,
\emph{J. Math. Phys.} \textbf{50} 095202 (2009).

\bibitem{PostWinternitz2015}
S.~Post and P.~Winternitz,
General $N$th order integrals of motion in the Euclidean plane,
\emph{J. Phys. A: Math. Theor.} \textbf{48} 405201 (2015).

\bibitem{MarquetteWinternitz2019}
I.~Marquette and P.~Winternitz,
Higher order superintegrability, Painlev\'e transcendents and representations of polynomial algebras,
\emph{J. Phys.: Conf. Ser.} \textbf{1194} 012074 (2019).

\bibitem{ELW2017}
A.~M.~Escobar-Ruiz, J.~C.~L\'opez Vieyra and P.~Winternitz,
Fourth order superintegrable systems separating in polar coordinates. I. Exotic potentials,
\emph{J. Phys. A: Math. Theor.} \textbf{50} 495206 (2017).

\bibitem{ELWCorrection}
A.~M.~Escobar-Ruiz, J.~C.~L\'opez Vieyra and P.~Winternitz,
Corrigendum: Fourth order superintegrable systems separating in polar coordinates. I. Exotic potentials,
\emph{J. Phys. A: Math. Theor.} \textbf{52} 019501 (2019).

\bibitem{MarquetteSajediWinternitz2017}
I.~Marquette, M.~Sajedi and P.~Winternitz,
Fourth order superintegrable systems separating in Cartesian coordinates I. Exotic quantum potentials,
\emph{J. Phys. A: Math. Theor.} \textbf{50} 315201 (2017).

\bibitem{CosgroveScoufis}
C.~M.~Cosgrove and G.~Scoufis,
Painlev\'e classification of a class of differential equations of the second order and second degree,
\emph{Stud. Appl. Math.} \textbf{88} 25--87 (1993).

\bibitem{TremblayWinternitz2010}
F.~Tremblay and P.~Winternitz,
Third-order superintegrable systems separating in polar coordinates,
\emph{J. Phys. A: Math. Theor.} \textbf{43} 175206 (2010).

\bibitem{ELVWY2018}
A.~M.~Escobar-Ruiz, J.~C.~L\'opez Vieyra, P.~Winternitz and \.I.~Yurdu\c{s}en,
Fourth-order superintegrable systems separating in polar coordinates. II. Standard potentials,
\emph{J. Phys. A: Math. Theor.} \textbf{51} 455202 (2018).

\bibitem{ERWY2018}
A.~M.~Escobar-Ruiz, P.~Winternitz and \.I.~Yurdu\c{s}en,
General $N$th-order superintegrable systems separating in polar coordinates,
\emph{J. Phys. A: Math. Theor.} \textbf{51} 40LT01 (2018).

\bibitem{GomezUllateKamranMilson2009}
D.~G\'omez-Ullate, N.~Kamran and R.~Milson,
An extended class of orthogonal polynomials defined by a Sturm--Liouville problem,
\emph{J. Math. Anal. Appl.} \textbf{359} 352--367 (2009).

\bibitem{Quesne2008}
C.~Quesne,
Exceptional orthogonal polynomials, exactly solvable potentials and supersymmetry,
\emph{J. Phys. A: Math. Theor.} \textbf{41} 392001 (2008).

\bibitem{MPR2020}
I.~Marquette, S.~Post and L.~Ritter,
A fourth-order superintegrable system with a rational potential related to Painlev\'e VI,
\emph{J. Phys. A: Math. Theor.} \textbf{53} 50LT01 (2020).

\bibitem{MPR2022}
I.~Marquette, S.~Post and L.~Ritter,
A family of fourth-order superintegrable systems with rational potentials related to Painlev\'e VI,
\emph{J. Phys. A: Math. Theor.} \textbf{55} 155201 (2022).

\bibitem{VinetZhedanovHeunHahn}
L.~Vinet and A.~Zhedanov,
The Heun operator of Hahn-type,
\emph{Proc. Amer. Math. Soc.} \textbf{147} 2987--2998 (2019).

\bibitem{BergeronEtAl}
G.~Bergeron, N.~Cramp\'e, S.~Tsujimoto, L.~Vinet and A.~Zhedanov,
The Heun--Racah and Heun--Bannai--Ito algebras,
\emph{J. Math. Phys.} \textbf{61} 081701 (2020).

\bibitem{KalninsMillerPost2007}
E.~G.~Kalnins, W.~Miller Jr and S.~Post,
Wilson polynomials and the generic superintegrable system on the 2-sphere,
\emph{J. Phys. A: Math. Theor.} \textbf{40} 11525--11538 (2007).

\bibitem{GenestVinetZhedanov2014}
V.~X.~Genest, L.~Vinet and A.~Zhedanov,
The Racah algebra and superintegrable models,
\emph{J. Phys.: Conf. Ser.} \textbf{512} 012011 (2014).

\bibitem{HGDR1984}
J.~Hietarinta, B.~Grammaticos, B.~Dorizzi and A.~Ramani,
Coupling-constant metamorphosis and duality between integrable Hamiltonian systems,
\emph{Phys. Rev. Lett.} \textbf{53} 1707--1710 (1984).

\bibitem{PostWinternitz2010}
S.~Post and P.~Winternitz,
An infinite family of superintegrable deformations of the Coulomb potential,
\emph{J. Phys. A: Math. Theor.} \textbf{43} 222001 (2010).

\bibitem{KalninsMillerPost2010}
E.~G.~Kalnins, W.~Miller Jr and S.~Post,
Coupling constant metamorphosis and $N$th-order symmetries in classical and quantum mechanics,
\emph{J. Phys. A: Math. Theor.} \textbf{43} 035202 (2010).

\bibitem{GrunbaumVinetZhedanov}
F.~A.~Gr\"unbaum, L.~Vinet and A.~Zhedanov,
Tridiagonalization and the Heun equation,
\emph{J. Math. Phys.} \textbf{58} 031703 (2017).

\bibitem{CheaVinetZhedanov2026}
V.~V.~Chea, L.~Vinet and A.~Zhedanov,
The 2D Smorodinsky--Winternitz II system and the Laguerre--Heun algebra,
\emph{J. Math. Phys.} \textbf{67} 102103 (2026).

\bibitem{JimboMiwa1981}
M.~Jimbo and T.~Miwa,
Monodromy preserving deformation of linear ordinary differential equations with rational coefficients. II,
\emph{Physica D} \textbf{2} 407--448 (1981).

\bibitem{OkamotoPVI}
K.~Okamoto,
Studies on the Painlev\'e equations I. Sixth Painlev\'e equation $P_{\rm VI}$,
\emph{Ann. Mat. Pura Appl.} \textbf{146} 337--381 (1987).

\bibitem{DLMF32}
P.~A.~Clarkson,
Painlev\'e transcendents, Chapter 32 of
\emph{NIST Digital Library of Mathematical Functions}, \url{https://dlmf.nist.gov/32}.

\bibitem{GIL2013}
O.~Gamayun, N.~Iorgov and O.~Lisovyy,
How instanton combinatorics solves Painlev\'e VI, V and IIIs,
\emph{J. Phys. A: Math. Theor.} \textbf{46} 335203 (2013).

\bibitem{DerezinskiRichard}
J.~Derezi\'nski and S.~Richard,
On Schr\"odinger operators with inverse square potentials on the half-line,
\emph{Ann. Henri Poincar\'e} \textbf{18} 869--928 (2017).

\end{thebibliography}
\end{document}